\documentclass[twocolumn]{fairmeta}

\usepackage[most]{tcolorbox}

\microtypesetup{expansion=false}
\usepackage{amsmath}
\usepackage{amssymb}

\title{RF-VoID: Towards Bandwidth-Efficient Exterior Tile Void Detection via Narrowband Radio-Frequency Representation Learning}

\author[1]{Xinyan Chen}
\author[2]{Ruiqin Ma}
\author[2]{Shunsuke Shoda}
\author[2]{Changyu Zhou}
\author[3]{Ryo Natsuaki}
\author[3]{Akira Hirose}
\author[1,*]{Jianfei Yang}
\author[2,3,*]{Li Yi}

\affiliation[1]{MARS Lab, Nanyang Technological University, Singapore}
\affiliation[2]{Ibaraki University, Mito, Ibaraki, Japan}
\affiliation[3]{The University of Tokyo, Bunkyo, Tokyo, Japan}

\contribution[*]{Corresponding author}

\abstract{
Hidden debonding behind exterior ceramic tiles is a falling-tile hazard, and millimeter-wave radar offers a non-contact way to find it. Conventional interpretation first reconstructs a range profile, so its reliability is bounded by the available bandwidth, yet bandwidth is what sets the cost, the acquisition time, and the regulatory footprint of a deployed system. This work asks whether that bandwidth can be traded for computation. A 4–40 GHz stepped-frequency system scans twelve exterior-wall specimens containing 0.5--1.0\,mm air voids at different depths and interfaces, and the bandwidth dependence of A-scan, B-scan, and C-scan interpretation is analyzed to establish the resolution bound. RF-VoID is then proposed, which decides directly on the narrowband complex response: the sub-band is kept in its measured frequency order with amplitude and phase alongside the in-phase and quadrature channels, a dual-branch encoder reads it along the physical frequency axis using relative position encoding and a distance-dependent locality bias, and an inspection-oriented objective handles the class imbalance and the asymmetric error cost of facade screening. Under a mixed-sample protocol the method attains 98.61\% accuracy and a 95.84\% F1-score with 0.5\,GHz of bandwidth, a seventy-two-fold reduction relative to the full sweep, without range-profile reconstruction, deconvolution, or depth-slice selection; on specimens held out entirely from training it remains the strongest of the compared models, with a mean macro F1-score of 62.12\% at 0.5 GHz that rises to 68.57\% at 1 GHz.
}

\correspondence{Jianfei Yang at \email{jianfei.yang@ntu.edu.sg}, Li Yi at \email{li.yi.wg60@vc.ibaraki.ac.jp}}
\metadata[Keywords]{millimeter-wave radar, exterior wall inspection, air void detection\\
narrowband, representation learning, nondestructive evaluation}

\begin{document}

\maketitle

\section{Introduction}\label{sec:intro}

Exterior ceramic-tile finishes are widely used in urban buildings because they provide durability, weather resistance, and architectural appearance. During long-term service, however, aging, thermal cycling, moisture ingress, construction defects, and seismic loading can weaken the bond among the tile, adhesive mortar, base mortar, and substrate. Hidden debonding or air voids may then form behind the tile layer, increasing the risk of falling tiles, especially on high-rise facades. Periodic inspection is therefore needed for public safety and preventive maintenance~\citep{Soeta2016}.

Conventional hammering or sounding inspection remains common in practice because it is simple and inexpensive. Recent studies have also explored data-driven interpretation of impact sounds~\citep{Ito2025}. Nevertheless, sounding inspection is labor intensive, requires close access to the facade, and is sensitive to inspector skill, site conditions, and environmental noise. Infrared thermography can support large-area screening of building finishes and adhesive tiling systems~\citep{Li2000,Lourenco2017}, but its reliability depends on thermal contrast and can be affected by solar loading, wind, emissivity variation, surface condition, and defect depth. These limitations motivate non-contact electromagnetic methods that can sense subsurface layer changes more directly.

Microwave and millimeter-wave nondestructive evaluation have been widely studied for layered structures, pavements, and interlayer debonding~\citep{Kharkovsky2007,Brinker2020,Yi2018}. Higher-frequency millimeter-wave radar systems can provide compact sensing configurations and improved depth sensitivity for shallow subsurface features~\citep{Chen2020,Koyabu2023}. Previous electromagnetic studies on exterior ceramic-tile inspection have also shown that hidden air voids can be visualized when sufficiently wide bandwidth is available~\citep{Alsalem2020,Shoda2025}. These studies also expose a critical limitation: conventional range-domain radar imaging becomes strongly bandwidth dependent when the void thickness is smaller than, or comparable to, the effective depth resolution of the measurement.

This dependence is what separates a laboratory demonstration from a deployable service, because wide bandwidth is considered costly for real-world applications. A 4--40\,GHz sweep needs a broadband source, a broadband antenna, and a vector network analyzer, whereas a sub-gigahertz measurement can in principle be made with a compact single-chip transceiver. Dwell time in a stepped-frequency acquisition is incurred per frequency point, so the points kept at each position set how long a facade scan takes and how much data it produces. Occupied bandwidth is also what active outdoor emission is regulated on, and what payload and power budgets squeeze hardest when the front end is pole- or drone-mounted~\citep{Koyabu2023}. Lowering the bandwidth an inspection needs is therefore not an accommodation of a hardware limit but the lever that makes low-cost, large-area screening feasible.
 
Two lines of response to bandwidth limitation already exist, and each fails in a way that matters here. The first tries to restore the missing resolution: sparse reconstruction~\citep{Yi2018}, deconvolution~\citep{Shoda2025}, and other super-resolution or compressed-sensing techniques~\citep{Wang2019,Candes2006,Donoho2006,Potter2010} can partially recover limited-bandwidth imaging, but their performance depends on accurate system-response estimation, sufficient signal-to-noise ratio, careful parameter tuning, and stable material properties, none of which is guaranteed on a weathered facade. The second applies learning to the reconstructed image, as CNN-based and deep-unfolding strategies have done across radar and remote-sensing tasks~\citep{Gao2018,Zhu2021}. Because the input to the network is a formed image, however, the resolution limit of the imaging stage propagates unchanged into the learned decision: a defect that is not separable in the reconstruction is not made separable by classifying that reconstruction. Bandwidth reduction degrades a learned decision made on images for exactly the same physical reason that it degrades a visual one.
 
The premise of this work is that bandwidth reduction destroys visibility, not information. A void of thickness $d$ enters the effective reflection coefficient of the wall through the round-trip factor $e^{-2\gamma d}$, whose argument is proportional to frequency, so the void perturbs amplitude and phase across the whole measured band rather than at one place in it. Conventional processing uses the inverse Fourier transform to concentrate that distributed perturbation into a range-domain peak, and only sufficient bandwidth lets the peak separate from its neighbors. A decision taken on the sub-band response itself never has to form the peak, and is therefore not bound by the resolution that forming it requires.
 
Acting on that premise takes more than a general-purpose classifier, because the setting imposes three demands. The input is an ordered complex response along a physical frequency axis, and since layer thickness and permittivity vary between walls, the same feature appears at a different absolute frequency in a different structure. A decision is needed at every scan position, where the void signature is a small perturbation of a strong, specimen-dependent background. And the labels are lopsided: sound wall dominates any scan, while a missed void is a safety failure and a false alarm costs one extra check. These three demands shape the input representation, the encoder, and the training objective respectively.
 
This work makes three contributions. First, a millimeter-wave dataset of exterior-wall specimens spanning different void depths, thicknesses, and interface locations is constructed, and used to quantify how conventional A-scan, B-scan, and C-scan interpretation degrades as bandwidth falls. Second, RF-VoID is proposed: a dual-branch encoder that acts directly along the frequency axis, on the measured sub-band response with amplitude and phase computed alongside its in-phase and quadrature channels. Its attention branch uses relative position encoding and a distance-dependent locality bias, so that the learned features transfer across specimens of differing layer geometry. Third, an inspection-oriented objective is formulated for the class imbalance and the asymmetric error cost of facade screening. The framework reaches 98.61\% accuracy and a 95.84\% overall F1-score at 0.5-GHz bandwidth under the mixed-sample protocol, supporting the claim that inspection bandwidth can be traded for computation.
 
The remainder of this paper is organized as follows. Section~\ref{sec:related} reviews microwave and millimeter-wave inspection of layered structures together with the learning-based methods that have been applied to it. Section~\ref{sec:imaging} analyzes the bandwidth dependence of conventional radar imaging on the present dataset. Section~\ref{sec:meas} describes the measurement system and the exterior-wall specimens. Section~\ref{sec:method} develops the proposed framework, Section~\ref{sec:results} reports and discusses the experimental results, and Section~\ref{sec:conclusion} concludes.

\section{Related Work}\label{sec:related}
 
\subsection{Microwave and Millimeter-Wave Inspection of Layered Civil Structures}\label{ssec:rw_nde}
 
Microwave and millimeter-wave nondestructive evaluation has a long record in civil and industrial inspection, and its physical basis is well established: a defect is detectable when it introduces an impedance discontinuity relative to the surrounding material~\citep{Kharkovsky2007,Brinker2020}. Effective-reflection models of layered dielectric slabs were used to detect disbonding and delamination well before high-frequency hardware became practical~\citep{Zoughi1990}, and the same principle underlies pavement and interlayer-debonding evaluation with ground-penetrating radar~\citep{Yi2018}. For exterior ceramic tiles specifically, microwave reflection measurements have been related to tile adhesion strength~\citep{Alsalem2020}, and wideband millimeter-wave measurements over 4--40\,GHz have been shown to visualize artificial voids behind the tile layer~\citep{Shoda2025}. Higher-frequency and array- or SAR-based systems continue to improve the achievable resolution of such measurements~\citep{Chen2020,Koyabu2023}.
 
Most of these studies share a two-stage inspection logic: the measurement is first converted into a range-domain or spatial image, and the decision is then made by reading that image. The quality of the decision is consequently bounded by the quality of the reconstruction, which for a thin defect is governed by the available bandwidth. It is also characteristic of this body of work that the reported detectability is obtained with laboratory-grade wideband instrumentation, and the limited bandwidth available in the field is not addressed. Section~\ref{sec:imaging} analyzes the bandwidth bound quantitatively on the present dataset; the remainder of this section reviews how learning-based methods have been introduced into the same inspection logic, and where they have departed from it.
 
\subsection{Learning-Based Interpretation of Radar Measurements}\label{ssec:rw_learning}
 
Most learning-based work in radar nondestructive evaluation replaces the human reading of the reconstructed image with a trained model, while leaving the imaging stage itself untouched. In ground-penetrating radar, convolutional detectors have been applied to B-scan images to locate reinforcement bars in concrete~\citep{Liu2020}, pattern-recognition workflows combining C-scan and B-scan evidence have been developed specifically for shallow subsurface air voids~\citep{Luo2020}, and deep-learning detectors have been used to identify and localize defects in urban underground space~\citep{Hu2023}. The same pattern appears at higher frequencies: delamination, debonding, and cavity defects in glass-fiber composite sandwich panels have been detected from terahertz images using an improved region-based convolutional network~\citep{Yang2023}. These methods are effective, but their inputs are formed images, whose resolution limitation largely affect the performance of the learning stage. A defect unresolved in the physical reconstruction cannot be resolved merely by classifying that image. Consequently, bandwidth reduction degrades learning-based detection for exactly the same physical reason that it impairs visual inspection.
 
A smaller body of work bypasses image formation altogether and learns from the measured signal. Near-field microwave reflection signals from a multilayered high-silica/phenolic composite have been classified with a one-dimensional convolutional network, with a classification-encoding scheme used to recover the two-dimensional extent of internal delaminations~\citep{Gao2025}. This establishes that a defect inside a layered material can be discriminated from the reflection signal itself, without an intervening reconstruction. The demonstration is nonetheless bounded by its measurement setting in ways that limit its reach as an inspection method. It requires near-field access to the specimen, a condition that a facade surveyed at stand-off distance cannot provide. It is carried out on a manufactured composite whose layer composition is controlled and repeatable, so the defect signature is separated from a background that is essentially the same in every sample, whereas a cast wall differs in layer thickness, material, and surface condition from one structure to the next. And detectability, rather than the cost of acquiring the signal, is the objective throughout, so how far the sensing bandwidth can be reduced before a signal-domain decision fails is left open.
 
\subsection{Learning Directly from Complex Radio-Frequency Responses}\label{ssec:rw_rf}
 
Outside nondestructive evaluation, a line of work in radio-frequency sensing has established that the raw complex response of a compact, often narrowband, radio carries enough information about a target's material composition for a learned model to exploit, with no imaging step at any point. Compact millimeter-wave radar has been used to categorize objects and materials placed on the sensor~\citep{Yeo2016}, and to drive fine-grained interaction from the sensor's radio-frequency signal rather than from a formed image~\citep{Wang2016}. With commodity hardware, material identification has been demonstrated from radio-frequency phase and amplitude measured through a target by passive tags~\citep{Wang2017,Xie2019}, from wideband channel measurements used to identify liquids inside containers~\citep{Dhekne2018}, and from the complex response of a single millimeter-wave radio~\citep{Wu2020}. Most recently, RF-MatID has assembled a dataset and benchmark for material identification directly from in-phase and quadrature responses~\citep{Chen2026}.
 
The recurring finding across this line of work is that dielectric contrast leaves a distinctive and learnable signature on the complex frequency response, and that recovering it does not require the bandwidth needed to resolve the target geometrically. The operating conditions, however, are consistently favorable. The target is a single homogeneous object, unobstructed or behind at most one thin barrier; the measurement geometry is controlled and repeatable; and the output is a single label per measurement, drawn from a closed set of classes that are represented in balanced proportion during training.
 
Exterior-wall inspection satisfies none of these conditions: the target is a sub-millimeter air layer buried inside a multilayer wall whose own interface reflections dominate the response, a decision is required at every position of a scan over a laterally varying structure rather than once per measurement, and the background itself differs from specimen to specimen. To the best of our knowledge, learning directly from narrowband complex radio-frequency responses has not previously been applied to point-wise hidden-void detection in multilayer building facades. Section~\ref{sec:method} develops a framework for this setting and Section~\ref{sec:results} evaluates it.

\section{Measurement System and Exterior-Wall Samples}\label{sec:meas}

\subsection{Millimeter-Wave Measurement System}\label{ssec:system}

\begin{figure}
  \centering
  \includegraphics[width=\columnwidth]{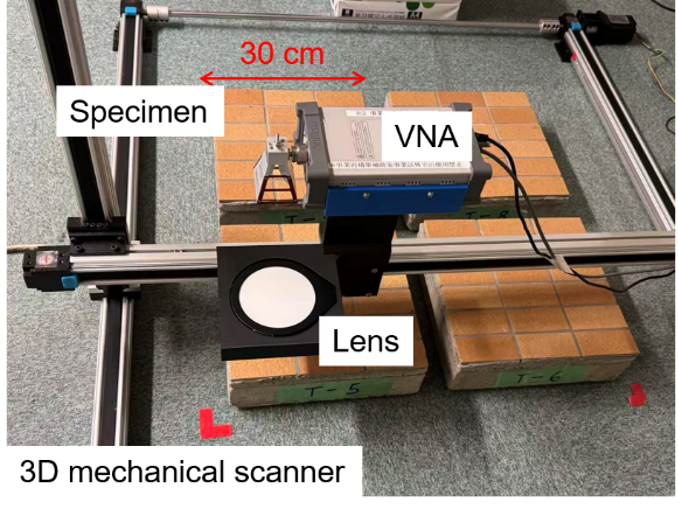}
  \vspace{-3mm}
  \caption{Millimeter-wave SFCW measurement system for exterior-wall inspection, including the vector network analyzer, horn antenna, Teflon lens, specimen, and three-dimensional mechanical scanner. }
  \label{fig:system}
\end{figure}

The measurement system follows a stepped-frequency continuous-wave (SFCW) radar configuration. It consists of a one-port vector network analyzer (Anritsu MS46131A), a broadband horn antenna (RF SPIN DRH40 double-ridged horn antenna), a Teflon lens, and a three-dimensional mechanical scanning stage, as shown in Fig.~\ref{fig:system}.

The emitted electromagnetic wave is normally incident on the specimen over the 4--40\,GHz frequency range. At each spatial sampling position, 2048 frequency points are recorded. In addition, for each spatial scan position $\mathbf{r}=(x,y)$, a narrow frequency sub-band can be extracted from the measured wideband SFCW response. The complex response at frequency $f_n$ is written as
\begin{equation}\label{eq:iq}
  S_{\mathbf{r}}(f_n) = I_{\mathbf{r}}(f_n) + jQ_{\mathbf{r}}(f_n),
\end{equation}
where $I_{\mathbf{r}}(f_n)$ and $Q_{\mathbf{r}}(f_n)$ denote the in-phase and quadrature components at spatial position $\mathbf{r}$ and frequency $f_n$, respectively. The amplitude
\begin{equation}\label{eq:amp}
  A_{\mathbf{r}}(f_n) = \lvert{S_{\mathbf{r}}(f_n)}\rvert=\sqrt{I_{\mathbf{r}}^{2}(f_n) + Q_{\mathbf{r}}^{2}(f_n)}
\end{equation}
reflects attenuation and scattering strength, whereas the phase
\begin{equation}\label{eq:phase}
  \phi_{\mathbf{r}}(f_n) = \operatorname{atan2}(Q_{\mathbf{r}}(f_n),I_{\mathbf{r}}(f_n))
\end{equation}
captures propagation delay and interference effects introduced by the layered wall structure. These amplitude- and phase-related variations provide the physical basis for detecting hidden air voids using millimeter-wave radar. Therefore, the complex I/Q response preserves defect-sensitive information that may remain useful even when a narrowband signal cannot be transformed into a clearly separated range-domain profile.

A Teflon lens with a focal length of 15\,cm and a diameter of 20\,cm is placed in front of the horn antenna to concentrate the radiated field on a localized region of the wall specimen. In the present setup, the lens-to-specimen distance is approximately 30\,cm, producing an estimated beam footprint of about 3\,cm in diameter.

This real-aperture configuration does not aim at extremely fine lateral resolution. Instead, it provides sufficient spatial localization for the present tile-void specimens, where practical debonding areas can extend over several centimeters and the main difficulty is depth-dependent interpretation within a multilayer wall. Another advantage of this configuration is the improved signal-to-noise ratio obtained by concentrating the local energy on the specimen. Compared with synthetic aperture radar (SAR) and array-based millimeter-wave imaging, which require accurate position information or multichannel calibration~\citep{Moreira2013,Sheen2001}, this real-aperture approach reduces acquisition and processing complexity.

\subsection{Exterior-Wall Samples}\label{ssec:specimens}

\begin{figure}
  \centering
  \includegraphics[width=\columnwidth]{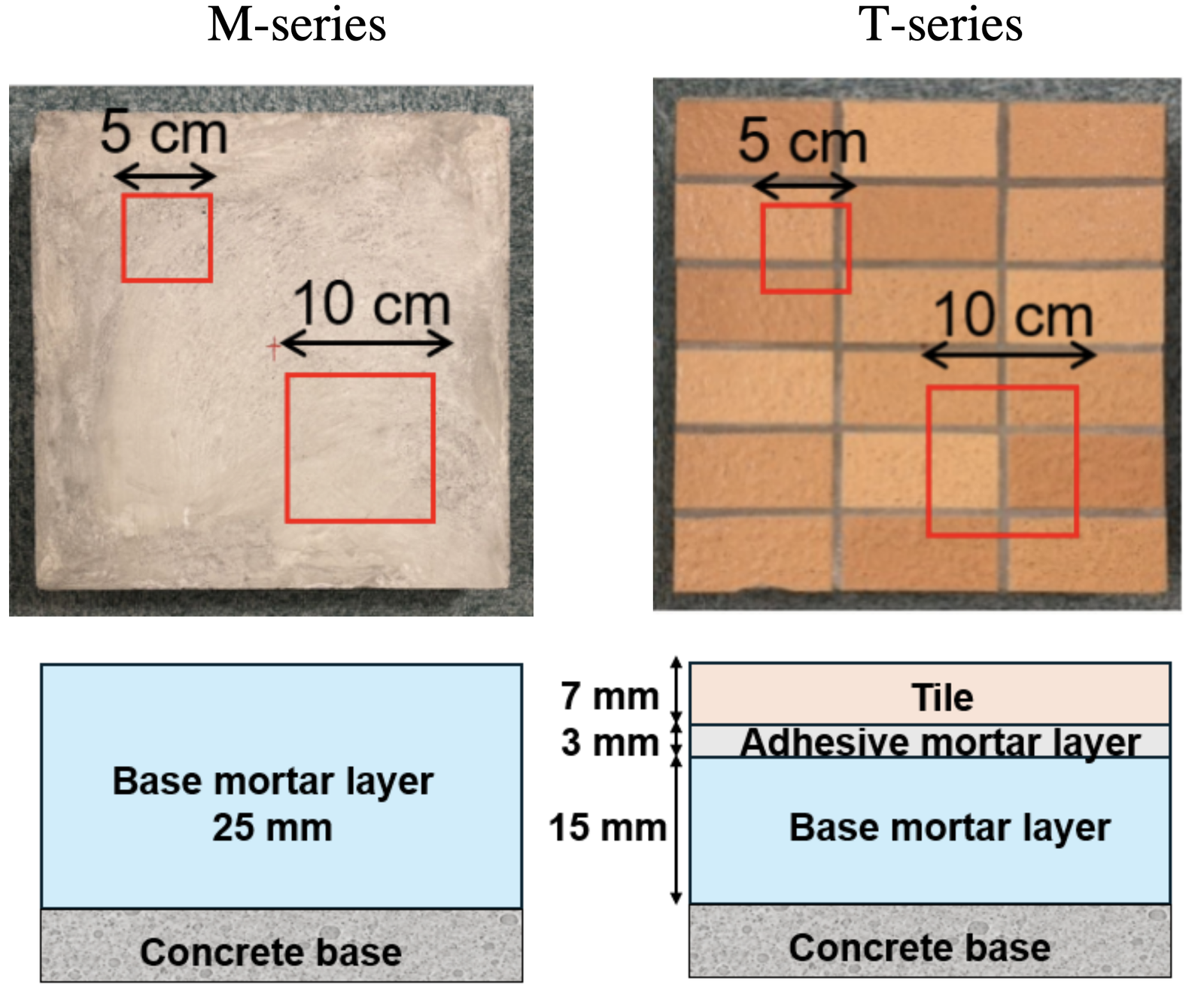}
  \vspace{-3mm}
  \caption{Representative exterior-wall specimens of T- and M-series and layer configuration used to generate controlled hidden air voids.}
  \label{fig:specimens}
\end{figure}

The specimens were designed to simulate exterior ceramic-tile wall structures containing hidden debonding or air-void defects. Each specimen consists of ceramic tile, adhesive mortar, base mortar, and concrete substrate layers. Thin inserts were placed during fabrication to form controlled air-void regions after curing. The artificial voids have lateral sizes of 50~$\times$~50\,mm and 100~$\times$~100\,mm, which are appropriate for studying practical facade debonding areas rather than extremely small lateral defects, as illustrated in Fig.~\ref{fig:specimens}.

Table~\ref{tbl:specimens} summarizes the specimen configurations. The set varies void depth, void thickness, interface location, and structural complexity. M-series specimens contain voids embedded in the base-mortar layer, whereas T-series specimens include voids at the tile/adhesive-mortar interface, the adhesive-mortar/base-mortar interface, the base-mortar/concrete interface, or two separated interfaces, as shown in Fig.~\ref{fig:specimens}. This design allows the radar response to be evaluated under progressively more difficult multilayer conditions.

From an electromagnetic perspective, an air void is detectable because it introduces an impedance discontinuity relative to the surrounding solid wall materials. This discontinuity changes the amplitude, phase, and multipath behavior of the reflected signal~\citep{Robert1998,Maierhofer2003}. However, the detectability of the void depends strongly on its depth, thickness, surrounding interfaces, and the available measurement bandwidth, which is further discussed in Section~\ref{sec:imaging}.

\begin{table}[tbp]
  \centering
\scalebox{0.74}{
    \begin{tabular}{c|l|c|c}
    \toprule
    \textbf{Specimen} & \multicolumn{1}{c|}{\textbf{Void Location}}                                   & \textbf{\begin{tabular}[c]{@{}c@{}}Void depth\\ (mm)\end{tabular}} & \textbf{\begin{tabular}[c]{@{}c@{}}Void Thickness\\ (mm)\end{tabular}} \\ \midrule
    M-1               & \multirow{4}{*}{\begin{tabular}[c]{@{}l@{}}Embedded within\\ base mortar\end{tabular}}             & 10                       & 0.5                          \\
    M-2               &                                                          & 10                       & 1.0                          \\
    M-3               &                                                          & 25                       & 0.5                          \\
    M-4               &                                                          & 25                       & 1.0                          \\ \midrule
    T-1               & \multirow{2}{*}{\begin{tabular}[c]{@{}l@{}}Tile/adhesive \\ mortar interface\end{tabular}}          & 7                        & 0.5                          \\
    T-2               &                                                          & 7                        & 1.0                          \\ \midrule
    T-3               & \multirow{2}{*}{\begin{tabular}[c]{@{}l@{}}Adhesive mortar /\\ base mortar interface\end{tabular}} & 10                       & 0.5                          \\
    T-4               &                                                          & 10                       & 1.0                          \\ \midrule
    T-5               & \multirow{2}{*}{\begin{tabular}[c]{@{}l@{}}base mortar/\\ concrete base\end{tabular}}              & 25                       & 0.5                          \\
    T-6               &                                                          & 25                       & 1.0                          \\ \midrule
    T-7               & \multirow{2}{*}{Two-interface voids}                     & 7/25                     & 0.5                          \\
    T-8               &                                                          & 7/25                     & 1.0                         \\ \bottomrule
    \end{tabular}
}
\caption{Specifications of the artificial void specimens. All specimens contained artificial voids with lateral sizes of $50 \times 50$\,mm and $100 \times 100$\,mm.}
\label{tbl:specimens}
\end{table}

Wideband measurements were acquired on a two-dimensional scanning grid with a spatial interval of 0.5 cm, as shown in Fig.~\ref{fig:system}. Each specimen is represented as a four-dimensional array of size 60 $\times$ 60 $\times$ 2048 $\times$ 2, where the first two dimensions denote spatial position, the third dimension denotes frequency sample, and the last dimension denotes the I/Q channels. Reduced-bandwidth subsets are extracted from the wideband measurements to emulate narrower radar operation under the same specimen geometry.

\section{Conventional Radar Imaging and Bandwidth Limitation}\label{sec:imaging}

\subsection{Thin-Void Detection Using A-Scan Signals}\label{ssec:ascan}

Conventional millimeter-wave inspection is generally performed by transforming the measured frequency-domain response into the range domain. At each scan position, the inverse Fourier transform of the complex frequency response provides an A-scan profile that represents reflected intensity along depth. For an SFCW measurement with bandwidth $B$, the nominal range resolution in air is given by
\begin{equation}\label{eq:res_air}
  \delta R = \frac{c}{2B},
\end{equation}
where $c$ is the speed of light. Inside the $i$-th dielectric layer, the propagation velocity is reduced by the square root of the relative permittivity $\varepsilon_{r,i}$, and the corresponding physical depth resolution can be approximated as
\begin{equation}\label{eq:res_layer}
  \delta z_{i} \approx \frac{c}{2B\sqrt{\varepsilon_{r,i}}}.
\end{equation}

For the full 4--40 GHz measurement, $B$ = 36 GHz, giving an air-range resolution of approximately 4.2 mm. Considering the relative permittivities $\varepsilon$ of ceramic tiles and mortars, which typically fall in the range of approximately 2--5~\citep{Robert1998, Maierhofer2003}, the corresponding physical depth resolution inside the wall layers is approximately 1.9--3.0 mm. This resolution is already larger than the 0.5--1.0 mm air-void thicknesses used in the specimens. Thus, even with the full measurement bandwidth, a thin void should not be expected to appear as two clearly separated boundary peaks. Reducing the bandwidth further broadens the range response in inverse proportion to $B$, thereby increasing the overlap among neighboring echoes.

The strength of each reflected component is governed by the impedance contrast. For the $i$-th layer, the wave impedance is
\begin{equation}\label{eq:impedance}
  \eta_{i} = \sqrt{\frac{\mu_{i}}{\varepsilon_{i}}},
\end{equation}
and the normal-incidence reflection coefficient between adjacent layers is
\begin{equation}\label{eq:refcoeff}
  \Gamma_{i,i+1} = \frac{\eta_{i+1} - \eta_{i}}{\eta_{i+1} + \eta_{i}}.
\end{equation}

For nonmagnetic construction materials, $\mu_{i}\approx\mu_{0}$; therefore, the impedance is mainly controlled by permittivity. Interfaces between solid materials with comparable dielectric properties, such as tile-to-mortar or mortar-to-concrete interfaces, generally produce moderate reflections. An air void introduces a stronger impedance discontinuity because $\varepsilon_{r,air}\approx1$. However, this does not guarantee that the void will be visually separable in a range-domain image.

In a multilayer wall, the measured response is a composite of surface reflection, internal material-interface reflections, void-boundary contributions, attenuation, phase delay, and multiple reflections~\citep{Maierhofer2003}. Using a standard transmission-line representation of layered dielectric media~\citep{Pozar2012}, the effective reflection coefficient can be written recursively as
\begin{equation}\label{eq:multilay}
  \Gamma^{\mathrm{eff}}_{i} = \frac{\Gamma_{i,i+1} + \Gamma^{\mathrm{eff}}_{i+1}
    \,e^{-2\gamma_{i+1} d_{i+1}}}{1 + \Gamma_{i,i+1}\,\Gamma^{\mathrm{eff}}_{i+1}
    \,e^{-2\gamma_{i+1} d_{i+1}}}
\end{equation}
Here, the complex propagation constant $\gamma_i$ of the corresponding layer depends on permittivity, permeability, conductivity, and frequency. It also accounts for both phase delay and attenuation. The layer-thickness term $d_i$ denotes the propagation distance through the layer. Similar effective-reflection models have been used for microwave detection of disbonding and delamination in layered dielectric slabs~\citep{Zoughi1990}. This model explains why a thin void may perturb the amplitude, phase, and multipath behavior of the measured response without necessarily appearing as an isolated range-domain peak.

\begin{figure}
  \centering
  \includegraphics[width=\linewidth]{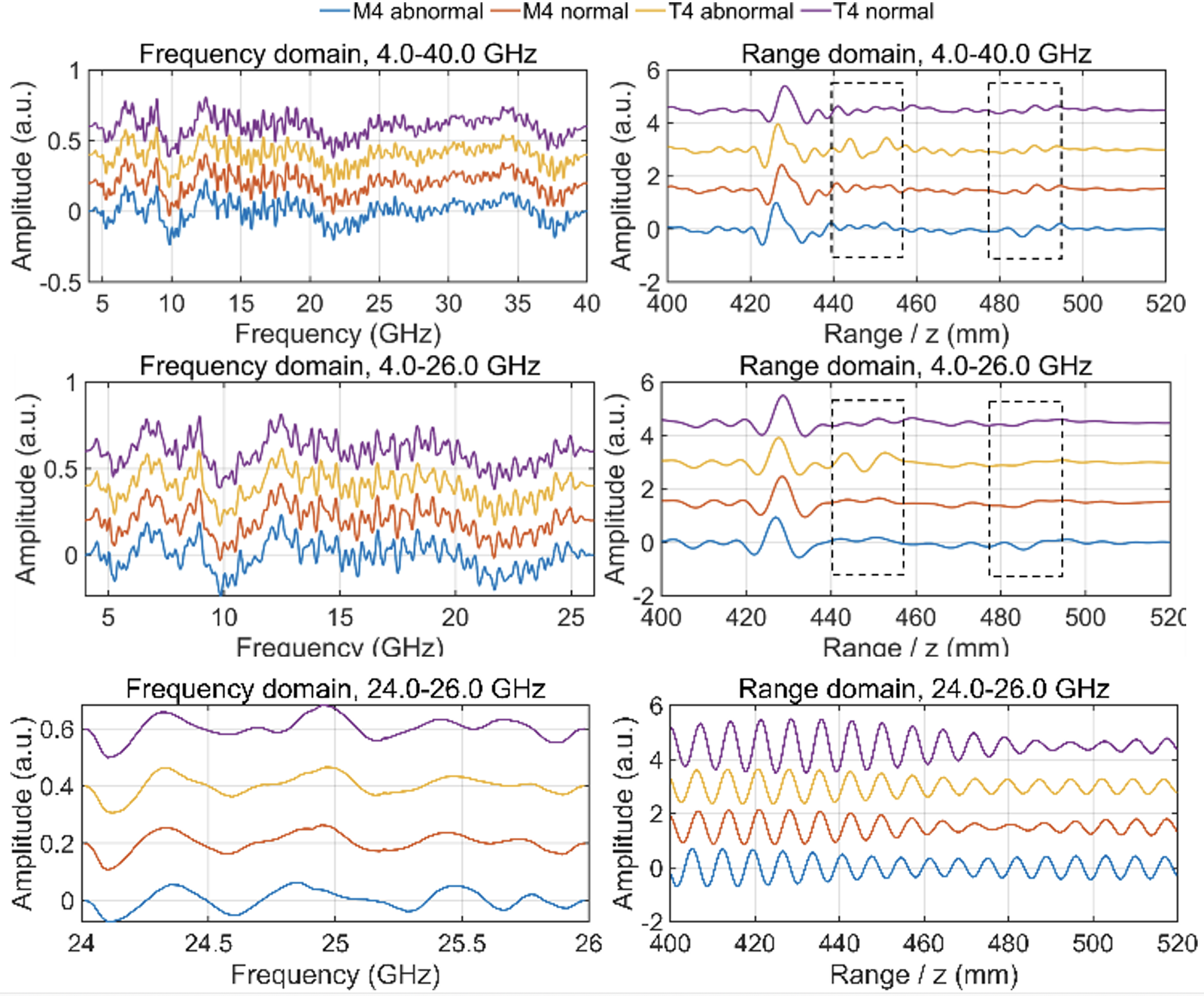}
  \caption{Representative A-scan responses of normal and void regions for T4 and M4 in the frequency and range domains under 4--40\,GHz, 4--26\,GHz, and 24--26\,GHz bandwidths. The dashed rectangles indicate the void responses at approximately 10 mm and 25 mm depths.}
  \label{fig:ascan}
\end{figure}

Fig.~\ref{fig:ascan} compares representative A-scan responses in the frequency and range domains. The strong response near 430\,mm corresponds to the specimen-surface response when the range axis is estimated using the propagation velocity in air. Under the full 4--40\,GHz bandwidth, the T4 abnormal trace shows a visible waveform perturbation near the expected deeper response region, whereas the M4 perturbation is weaker and occurs farther in range, around approximately 480\,mm. The 4--26\,GHz result remains qualitatively close to the full-band result, indicating that much of the useful penetration energy is retained in this band. In the narrow 24--26\,GHz case, ringing and peak broadening become pronounced, making the void-related perturbation difficult to isolate from surface and layer-interface responses.

Notably, these waveform variations are also affected by stand-off distance, incidence angle, layer-thickness variation, material nonuniformity, and multilayer structural complexity. Consequently, direct visual interpretation of A-scan responses alone is not sufficiently robust for practical inspection. This motivates the subsequent use of spatial imaging and AI-based interpretation.

\subsection{Imaging-Based Inspection and Its Limitations}\label{ssec:imaging}

Compared with direct interpretation of A-scan waveforms, B-scan and C-scan images provide more intuitive spatial views for radar-based nondestructive inspection~\citep{Kharkovsky2007,Brinker2020,Wang2019}. By stacking A-scans along a scan line, a B-scan cross section is obtained. By extracting the response at a selected depth over the two-dimensional scanning grid, a C-scan image is generated. These representations are familiar and physically interpretable because they link the measured RF response to layer depth and lateral position.

The representative specimens T4, M4, and T7 were selected to illustrate progressively more difficult imaging conditions. T4 contains a 1.0\,mm void at the adhesive-mortar/base-mortar interface and represents a relatively simple single-interface case. M4 contains a deeper 1.0\,mm void embedded in the base mortar, so its response is more attenuated and more likely to overlap with shallower echoes. T7 contains two 0.5\,mm voids at different interfaces, producing coupled contributions from multiple air voids and neighboring material boundaries.

\begin{figure}
  \centering
  \includegraphics[width=\columnwidth]{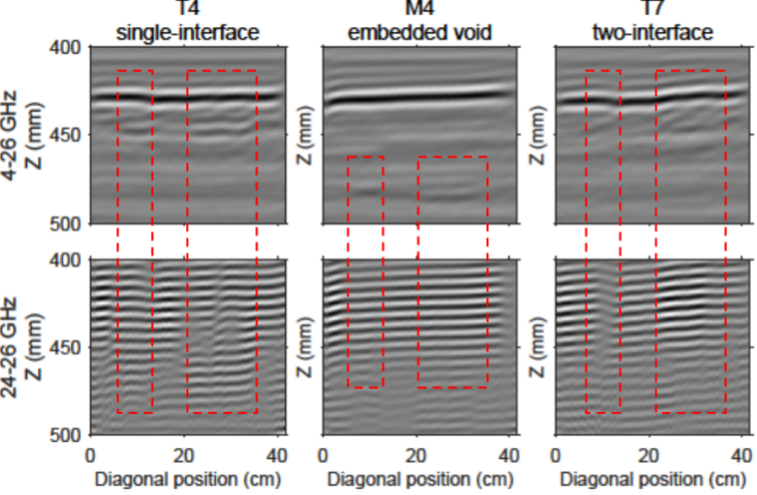}
  \vspace{-3mm}
  \caption{Representative diagonal B-scan images for T4, M4, and T7 under wider-band (4--26\,GHz) and narrower-band (24--26\,GHz) conditions.}
  \label{fig:bscan}
\end{figure}

Fig.~\ref{fig:bscan} illustrates the imaging results using B-scan images. The diagonal scan crosses both void regions and provides a depth-versus-position view of the reflected response. When the expected void locations are known, wider-band B-scans can be interpreted in relation to local horizontal anomalies. Without prior knowledge, however, these anomalies can be confused with surface and internal-boundary reflections. Under narrower bandwidths, the B-scan signatures broaden and become less separable along depth, weakening the visual connection between the observed response and the true void location. Notably, the 4--26\,GHz bandwidth still provides clear recognition of the void regions. A 4--18\,GHz case, examined on the same specimens but omitted from Fig.~\ref{fig:bscan} for brevity, lies close to the lower limit at which the anomaly can still be identified visually. This is because the void-related reflections become increasingly difficult to separate from the surface reflection and neighboring layer-interface responses as the bandwidth decreases. A deconvolution approach can be applied to B-scan images to enhance the range resolution by estimating the source wavelet, and~\citet{Shoda2025} report that void-related responses can then be partially restored under a 4--11\,GHz bandwidth. However, the restored results are strongly dependent on parameter selection, source-wavelet estimation, signal-to-noise ratio, and sample conditions. Therefore, applying such methods robustly across the entire specimen dataset remains challenging, which is a known limitation of conventional deconvolution and sparse-reconstruction-based approaches.

\begin{figure}
  \centering
  \includegraphics[width=\columnwidth]{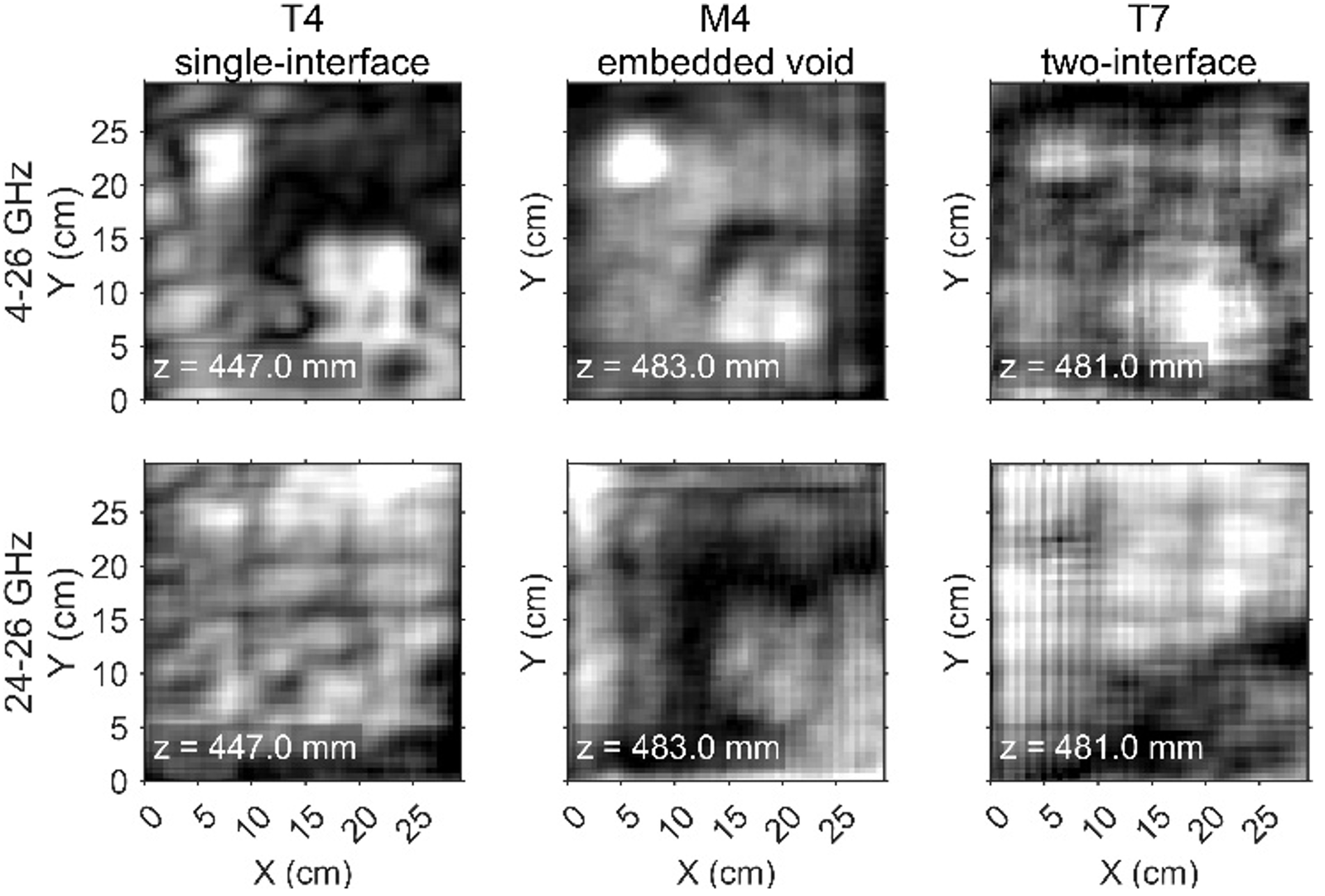}
  \vspace{-3mm}
  \caption{Representative C-scan images for T4, M4, and T7 under wider-band (4--26\,GHz) and narrower-band (24--26\,GHz) conditions.}
  \label{fig:cscan}
\end{figure}

Fig.~\ref{fig:cscan} shows C-scan images for T4, M4, and T7 extracted at representative depth slices selected according to the void-related response regions observed in the A-scan and B-scan results. C-scan visualization can highlight spatially continuous anomalies, but its reliability depends on choosing an appropriate depth slice. Under wider-band conditions, the void region is more recognizable, particularly for the simpler T4 case. As bandwidth decreases and structural complexity increases, the selected slice contains more overlapped contributions. For M4, the deeper void remains only weakly visible in some narrowband cases, whereas in T7 the surface-related shadow and reduced range resolution can obscure the void response.

When the full 4--40\,GHz bandwidth is used, the reconstructed radar signals can reveal the main internal structure of the exterior-wall specimens, as illustrated in Figs.~\ref{fig:ascan}--\ref{fig:cscan}. However, even under controlled laboratory fabrication, void-induced responses are not always ideal or perfectly shaped. Local variations in mortar thickness, surface condition, material nonuniformity, and imperfect void formation can distort the reflected signals. In such cases, C-scan imaging is useful because spatial continuity across neighboring scan positions can make void-related anomalies easier to distinguish from local clutter.

This advantage, however, comes at the cost of dense two-dimensional data acquisition and careful depth selection. Reliable C-scan imaging requires the specimen surface to be scanned with a sufficiently small spatial interval, typically determined by the wavelength, beam footprint, or desired lateral resolution. Moreover, the displayed result depends on the selected depth slice: if the extraction depth is slightly mismatched, the void contrast may weaken substantially. Although array-based and SAR-based millimeter-wave imaging has been investigated for high-resolution measurement~\citep{Chen2020,Koyabu2023,Wang2019}, its hardware complexity, calibration burden, data volume, and computational cost remain considerably higher than those of simpler narrowband sensing configurations.

\section{AI-Based Narrowband RF Representation Learning}\label{sec:method}

\begin{figure*}
  \centering
  \includegraphics[width=\textwidth]{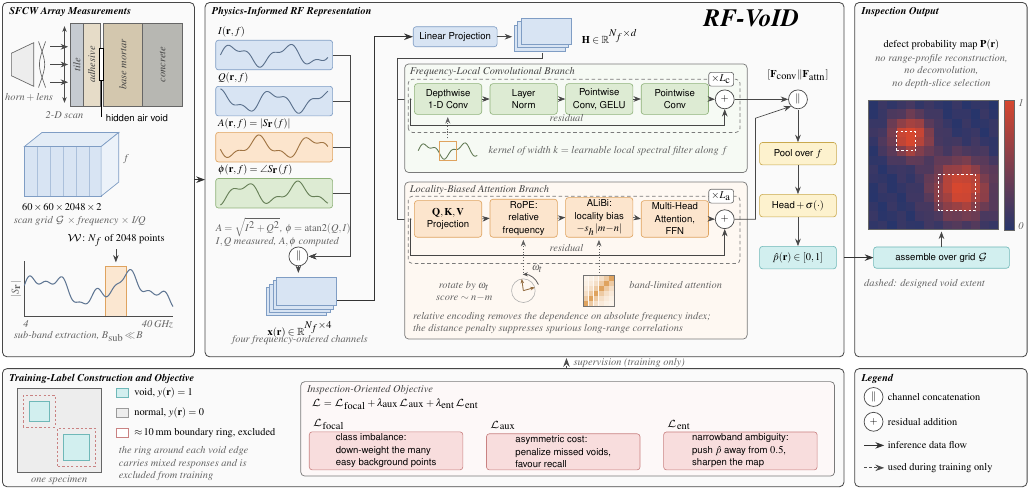}
  \caption{Overview of the proposed narrowband RF representation learning framework. Each dashed block is the repeated unit of its branch, applied $L_{\mathrm{c}}$ and $L_{\mathrm{a}}$ times respectively; the dotted call-outs below the branches illustrate the core idea of the module they point to. The wideband SFCW response measured at each scan position is restricted to a narrow sub-band, represented by four physically interpretable channels, encoded by the dual-branch RF-VoID, and converted into a point-wise defect probability map over the scanning grid. Training labels are constructed from the designed void geometry, with the ambiguous boundary ring around each void edge excluded from supervision.}
  \label{fig:pipeline}
\end{figure*}

Consider a single scan position $\mathbf{r}=(x,y)$ on the specimen surface, at which the radar records the complex response $S_{\mathbf{r}}(f_n)$ of~\eqref{eq:iq} over a narrow sub-band selected from the full 4--40\,GHz sweep. The sub-band is the set $\mathcal{W}$ of $N_f$ consecutive measured frequency points, written $f_1,\ldots,f_{N_f}$ after re-indexing within $\mathcal{W}$, and its width $B_{\mathrm{sub}}=f_{N_f}-f_1$ is far smaller than the full measurement bandwidth $B=36$\,GHz. Throughout, $\mathcal{W}$ denotes the frequency-sampling set in the same way that $\mathcal{G}$ denotes the spatial-sampling set. Let $y(\mathbf{r})\in\{0,1\}$ indicate whether a hidden air void lies beneath $\mathbf{r}$. The inspection task is to estimate the void posterior $\hat{p}(\mathbf{r})=\Pr\{y(\mathbf{r})=1\mid S_{\mathbf{r}}(\mathcal{W})\}$. Conventional inspection reaches this decision indirectly: the sub-band response is inverted into a range profile, a depth slice is selected, and the decision is made from the reflected amplitude at that depth. Section~\ref{sec:imaging} showed that every link of this chain weakens as $B_{\mathrm{sub}}$ decreases, because the depth resolution $\delta z$ of~\eqref{eq:res_layer} grows beyond the void thickness and the void echo can no longer be isolated. This work therefore replaces the chain with a single learned mapping
\begin{equation}\label{eq:task}
  \hat{p}(\mathbf{r}) = \mathcal{F}_{\theta}\bigl(\mathbf{x}(\mathbf{r})\bigr),
  \qquad \hat{p}(\mathbf{r})\in[0,1],
\end{equation}
in which $\mathbf{x}(\mathbf{r})$ is a representation of the sub-band response measured at $\mathbf{r}$ and the parameters $\theta$ are obtained by minimizing an inspection-oriented loss over labeled scan positions. Neither range-profile reconstruction, nor deconvolution, nor depth-slice selection appears in~\eqref{eq:task}.

As illustrated in Fig.~\ref{fig:pipeline}, $\mathcal{F}_{\theta}$ is realized in three stages. First, the selected sub-band is converted into a physics-informed multi-channel sequence that retains the measured complex response together with the amplitude and phase quantities that carry its physical meaning. Second, the sequence is encoded by two complementary branches that operate directly along the frequency axis: a frequency-local convolutional branch that responds to short-scale spectral fluctuation, and a locality-biased attention branch that relates frequency samples separated by larger intervals. Third, the two branch outputs are fused and mapped by a lightweight head to the void posterior of~\eqref{eq:task}, which is written back to the scanning grid to form a defect probability map. The resulting model is denoted RF-VoID. In it's design, three task-specific modifications are introduced, and all of them follow one principle: the index axis of the input is a physical frequency axis, so every operation applied to it should preserve that ordering rather than discard it.

The remainder of this section develops $\mathcal{F}_{\theta}$ in the order of Fig.~\ref{fig:pipeline}. Section~\ref{ssec:motivation} explains, from the layered-medium model of Section~\ref{sec:imaging}, why void-related information survives bandwidth reduction even when it is no longer visible in the range domain. Section~\ref{ssec:iq} constructs the input representation, Section~\ref{ssec:model} describes the dual-branch encoder, Section~\ref{ssec:loss} defines the training objective, and Section~\ref{ssec:probmap} generates the defect probability map.

\subsection{From Range-Domain Imaging to Learned RF Representation}\label{ssec:motivation}

The analysis of Section~\ref{sec:imaging} is easily read as a statement about information: if a thin void cannot be resolved in the range profile, the narrowband measurement is assumed to be insufficient. The multilayer reflection model of~\eqref{eq:multilay} suggests a different reading. A void of thickness $d$ enters the effective reflection coefficient through the round-trip factor $e^{-2\gamma d}$, whose argument is proportional to frequency. The void therefore does not contribute at one isolated frequency; it modulates the amplitude and the phase of the response across the whole measured band. The inverse Fourier transform concentrates this distributed modulation into a compact range-domain feature, and it does so successfully only when the bandwidth is large enough that $\delta z$ in~\eqref{eq:res_layer} is smaller than the separation between the void boundaries and the neighboring interfaces. When this condition fails, the modulation is not removed from the measurement; it is spread over the sub-band and superimposed on the much stronger surface and interface responses.

This distinction motivates the approach taken here. Rather than attempting to undo the bandwidth limitation through super-resolution, deconvolution, or sparse reconstruction, which require an accurate estimate of the system wavelet, a favorable signal-to-noise ratio, and case-dependent parameter tuning~\citep{Yi2018,Shoda2025,Wang2019}, the decision is made on the sub-band response itself. The task is thus reformulated from resolving the void to discriminating the spectral signature that its presence imposes on the measured response.

Learning directly from complex frequency-domain data has precedent in RF sensing. mSense~\citep{Wu2020} and RF-MatID~\citep{Chen2026} show that a network can infer material properties from in-phase and quadrature responses without any conversion to the range domain. The present task differs from those in two respects that shape the model design. First, material identification assigns one label to a homogeneous sample placed in front of the antenna, whereas exterior-wall inspection requires a decision at every scan position of a laterally varying structure. Second, and more restrictively, the quantity to be detected here is a sub-millimeter air layer buried inside a multilayer stack whose own reflections dominate the response, so the discriminative pattern is a small perturbation of a strong and specimen-dependent background rather than a distinctive signature of the target itself. These two differences motivate the physics-informed input of Section~\ref{ssec:iq} and the frequency-aware encoder of Section~\ref{ssec:model}.

\subsection{Physics-Informed Narrowband RF Representation}\label{ssec:iq}

As illustrated in Fig.~\ref{fig:pipeline}, the measured wideband SFCW response is first restricted to a selected narrow sub-band and then represented by its in-phase and quadrature components at each spatial scan position. Since each frequency-domain response is written as~\eqref{eq:iq}, the input sequence for the $N_f$ frequency samples within the sub-band is represented as
\begin{equation}
  \begin{split}
    \mathbf{x}^{\mathrm{IQ}} = \bigl[&I(\mathbf{r}, f_1),\, Q(\mathbf{r}, f_1),\,
      \ldots,\\
      &I(\mathbf{r}, f_{N_f}),\, Q(\mathbf{r}, f_{N_f})\bigr]
      \in \mathbb{R}^{N_f\times2}
  \end{split}
\end{equation}

This representation preserves the ordered complex frequency response without explicitly converting it into a range-domain profile. Compared with conventional radar signal processing~\citep{Richards2014}, which relies on separable reflection peaks after inverse Fourier transformation, the proposed representation allows the model to use subtle amplitude, phase, and interference patterns that may remain in the narrowband RF signal.

To provide additional physically meaningful information, we further obtain the magnitude and phase features:
\begin{align}
  \mathbf{x}^{A}   &= \bigl[A(\mathbf{r}, f_1),\, \ldots,\,
    A(\mathbf{r}, f_{N_f})\bigr], \\
  \mathbf{x}^{\phi} &= \bigl[\phi(\mathbf{r}, f_1),\, \ldots,\,
    \phi(\mathbf{r}, f_{N_f})\bigr].
\end{align}

The augmented input is then given by
\begin{equation}\label{eq:input}
  \mathbf{x} = \bigl[\mathbf{x}^{\mathrm{I}},\; \mathbf{x}^{\mathrm{Q}},\; \mathbf{x}^{A},\; \mathbf{x}^{\phi}\bigr] \in \mathbb{R}^{N_f\times4}.
\end{equation}
Though the original I/Q channels basically retain the complex-valued measurement information, the values completely depend on the reference phase in the observatoin. In contrast, the magnitude and phase channels provide explicit cues related to reflection strength and relative propagation delay. In particular, the phase value is highly informative to represent the RF wave physics \citep{konishi2023insar}.

\subsection{Dual-Branch Narrowband RF Encoder}\label{ssec:model}

The representation of~\eqref{eq:input} is a sequence indexed by physical frequency, and two questions about that axis decide what should act on it: over what range along the axis does a void act, and what does a position on the axis mean.

The first question separates two scales. A void does not perturb one part of the sub-band in particular: the extra round-trip delay and the modified impedance contrast it introduces change the slope and the curvature of amplitude and phase across the whole of $\mathcal{W}$, which is a relation between frequency samples that lie far apart. Superimposed on that are fluctuations that change from one sample to the next, produced by multiple reflections between closely spaced interfaces and by measurement noise. An operator sized for one of these scales is mismatched to the other, so the encoder is split into a branch that reads the response over a short window of adjacent samples and a branch that relates samples across the band.

The second question decides how position is encoded. In a layered medium the interference between interface echoes is periodic in frequency, with a period $c/(2d\sqrt{\varepsilon_r})$ governed by the thickness $d$ and permittivity of the layer rather than by where in the spectrum the sub-band happens to be taken. Depth information is therefore carried by the separation between two frequency samples and not by their absolute indices, and since layer thickness and permittivity differ between specimens, the same physical feature sits at a different absolute index in a different wall. That coupling is also not uniform in separation: over a narrow sub-band the per-sample signal-to-noise ratio limits how far a genuine correlation can be traced before broadband system residuals dominate. Position must consequently enter the model through differences, and those differences must be weighted by how large they are, which is what the relative encoding and the locality bias of the attention branch supply.

Both branches operate on a shared linear projection
\begin{equation}\label{eq:proj}
  \mathbf{H} = \mathrm{Proj}\bigl(\mathbf{x}(\mathbf{r})\bigr) \in \mathbb{R}^{N_f\times D},
\end{equation}
where $\mathrm{Proj}(\cdot)$ maps the four physical channels to a hidden dimension $D$ at each frequency sample and leaves the length of the sequence, and therefore the frequency ordering, unchanged.

\textbf{Frequency-local convolutional branch.} The original RF-MatID folds the one-dimensional response into a two-dimensional map so that image backbones can be reused~\citep{Chen2026}. Folding places frequency samples that are far apart in the spectrum into adjacent rows, so a two-dimensional kernel mixes spectrally unrelated samples, and the dense layers that follow introduce parameters that are difficult to constrain when $N_f$ is small. RF-VoID removes the folding operation and adopts a native one-dimensional ConvNeXt structure~\citep{Liu2022},
\begin{equation}\label{eq:conv}
  \mathbf{F}_{\mathrm{conv}} = \mathrm{ConvNeXt1D}(\mathbf{H};\theta_{\mathrm{conv}}),
\end{equation}
in which depthwise kernels slide along the physical frequency axis. The block of~\eqref{eq:conv}, comprising a depthwise convolution, normalization, and two pointwise convolutions with a residual connection, is repeated $L_{\mathrm{c}}$ times. Each kernel then acts as a learnable local spectral filter over a window of adjacent frequency samples, which is the operation a conventional signal-processing treatment would apply to a frequency response, and the topology of the spectrum is preserved rather than rearranged.

\textbf{Locality-biased attention branch.} Band-wide dependence is modeled by self-attention~\citep{Vaswani2017}, which computes
\begin{equation}\label{eq:qkv}
  \mathbf{Q} = \mathbf{H}\mathbf{W}_Q,\quad
  \mathbf{K} = \mathbf{H}\mathbf{W}_K,\quad
  \mathbf{V} = \mathbf{H}\mathbf{W}_V,
\end{equation}
and weights each pair of frequency samples by the similarity between a query and a key. Attention is by construction insensitive to ordering, and the usual remedy of adding an absolute positional encoding is precisely what the second consideration above rules out, since it ties every learned pattern to a fixed frequency index. Rotary position embedding (RoPE)~\citep{Su2024} is used instead. Query and key vectors at frequency index $m$ are rotated by an angle proportional to $m$,
\begin{equation}\label{eq:rope}
  \mathbf{R}_{m}\,\mathbf{q}_m =
  \begin{pmatrix} \cos m\omega_t & -\sin m\omega_t \\
                  \sin m\omega_t & \phantom{-}\cos m\omega_t \end{pmatrix}
  \begin{pmatrix} q_{m,2t-1} \\ q_{m,2t} \end{pmatrix},
\end{equation}
applied to each pair $t=1,\ldots,D/2$ of hidden dimensions with a pair-specific angular frequency $\omega_t$, so that the inner product between a rotated query at index $m$ and a rotated key at index $n$ satisfies
\begin{equation}\label{eq:rope_rel}
  \bigl\langle \mathbf{R}_{m}\mathbf{q}_m,\;
  \mathbf{R}_{n}\mathbf{k}_n \bigr\rangle
  = \mathbf{q}_m^{\top}\,\mathbf{R}_{n-m}\,\mathbf{k}_n .
\end{equation}
The interaction thus depends on the frequency separation $n-m$ rather than on absolute positions, which is the property the periodicity of the layered response calls for.

Relative encoding alone, however, still lets every pair of frequency samples interact equally, and the reliability of a correlation falls off with separation for the reason given above. Attention with linear biases (ALiBi)~\citep{Press2022} is therefore applied, which subtracts a penalty proportional to the frequency-index distance before the softmax operation,
\begin{equation}\label{eq:alibi}
  \mathrm{A}_{mn} = \frac{\mathbf{q}_m^{\top}\,\mathbf{R}_{n-m}\,\mathbf{k}_n}
  {\sqrt{D_k}} - s_h\,\lvert m-n\rvert,
\end{equation}
where the first term is the rotary score of~\eqref{eq:rope_rel} scaled by the per-head key dimension $D_k$, $\lvert m-n\rvert$ is the index distance between two frequency samples, and $s_h$ is a head-specific slope. The penalty expresses the prior that adjacent frequency samples are more strongly coupled than distant ones. Because $s_h$ differs across heads, the branch inspects the sub-band at several effective widths at once, from a narrow neighborhood around each sample to almost the entire band, so that local spectral changes and band-wide trends are represented within the same branch. The attention block, comprising~\eqref{eq:qkv}--\eqref{eq:alibi} followed by a feed-forward stage and a residual connection, is repeated $L_{\mathrm{a}}$ times, and its output is denoted $\mathbf{F}_{\mathrm{attn}}$.

\textbf{Branch fusion and inspection head.} The two branch outputs are combined by a learnable weighting, concatenated along the channel dimension, pooled over the frequency axis, and mapped to a scalar by a lightweight classification head,
\begin{equation}\label{eq:head}
  \hat{p}(\mathbf{r}) = \sigma\Bigl(\mathrm{Head}\bigl(
    [\mathbf{F}_{\mathrm{conv}}\,\Vert\,\mathbf{F}_{\mathrm{attn}}]\bigr)\Bigr),
\end{equation}
where $[\,\cdot\,\Vert\,\cdot\,]$ denotes concatenation and $\sigma(\cdot)$ is the sigmoid function. Equations~\eqref{eq:proj}--\eqref{eq:head} together realize the mapping $\mathcal{F}_{\theta}$ of~\eqref{eq:task}.

\subsection{Inspection-Oriented Training Objective}\label{ssec:loss}

Let $y(\mathbf{r})\in\{0,1\}$ denote the ground-truth label at scan position $\mathbf{r}$, where $y=1$ indicates a void point and $y=0$ a normal point; label construction and the exclusion of ambiguous boundary positions are described in Section~\ref{ssec:protocols}. Training $\mathcal{F}_{\theta}$ on these labels with a standard cross-entropy objective is inadequate for three reasons that are specific to narrowband inspection.

First, the label distribution is strongly imbalanced. Void regions occupy a small fraction of each scanned specimen, and normal positions outnumber void positions by roughly an order of magnitude, so an objective that weights all positions equally is dominated by background points that are already easy to classify. Second, the two error types are not equally costly. In falling-tile prevention a missed void is a safety failure, whereas a false alarm leads only to an additional manual check, so recall must be prioritized over a symmetric error rate. Third, the narrowband responses of void and normal positions are close to each other, and the model consequently produces many predictions near $\hat{p}=0.5$; once these are thresholded, they generate scattered false detections that degrade precision and break the spatial continuity of the resulting map.

The training objective addresses the three effects explicitly and is written as
\begin{equation}\label{eq:loss}
  \mathcal{L} = \mathcal{L}_{\mathrm{focal}}
  + \lambda_{\mathrm{aux}}\,\mathcal{L}_{\mathrm{aux}}
  + \lambda_{\mathrm{ent}}\,\mathcal{L}_{\mathrm{ent}},
\end{equation}
where $\lambda_{\mathrm{aux}}$ and $\lambda_{\mathrm{ent}}$ balance the three terms.

The focal term $\mathcal{L}_{\mathrm{focal}}$~\citep{Lin2017} handles the imbalance. It rescales the cross-entropy of each position by a factor that decreases as the prediction becomes confident and correct, so the gradient contributed by the many easy background positions is suppressed and the optimization is directed towards positions that remain misclassified.

The auxiliary anomaly term addresses the asymmetric cost. It is evaluated only on the $N_{\mathrm{void}}$ labeled void positions,
\begin{equation}\label{eq:aux}
  \mathcal{L}_{\mathrm{aux}} = -\frac{1}{N_{\mathrm{void}}}
  \sum_{\mathbf{r}:\,y(\mathbf{r})=1} \log \hat{p}(\mathbf{r}),
\end{equation}
and therefore penalizes low predicted probability on true voids without being offset by the far more numerous normal positions. It acts as a direct penalty on missed detections and consequently favors recall.

The entropy term addresses the ambiguity of narrowband responses,
\begin{equation}\label{eq:ent}
  \begin{split}
    \mathcal{L}_{\mathrm{ent}} = -\frac{1}{N}\sum_{\mathbf{r}}
    \bigl[&\hat{p}(\mathbf{r})\log(\hat{p}(\mathbf{r})+\epsilon)\\
    &+ (1-\hat{p}(\mathbf{r}))\log(1-\hat{p}(\mathbf{r})+\epsilon)\bigr],
  \end{split}
\end{equation}
where the sum runs over the $N$ training positions and $\epsilon$ is a small constant added for numerical stability. The expression is the mean binary entropy of the predictions and is maximized at $\hat{p}=0.5$, so minimizing it moves the model away from indecisive outputs. The effect is not only numerical: sharper per-position decisions also produce more clearly delimited void regions. The weight $\lambda_{\mathrm{ent}}$ is kept small, so that this term regularizes the decision without overriding the supervised terms.

The three terms act on different aspects of the same difficulty. The focal term determines which positions the optimization attends to, the auxiliary term determines how the two error types are traded off, and the entropy term determines how decisively the model is required to commit at each position.

\subsection{Defect Probability Map Generation}\label{ssec:probmap}

After training, $\mathcal{F}_{\theta}$ is applied independently at every position of the two-dimensional scanning grid $\mathcal{G}$ described in Section~\ref{ssec:specimens}, and the outputs are written back to the measurement coordinates to form a defect probability map
\begin{equation}\label{eq:probmap}
  \mathbf{P}(\mathbf{r}) = \mathcal{F}_{\theta}\bigl(\mathbf{x}(\mathbf{r})\bigr),
  \qquad \mathbf{r}\in\mathcal{G}.
\end{equation}

The map resembles a C-scan image, but the quantity displayed at each pixel is different, and that difference is what removes the depth-selection problem identified in Section~\ref{ssec:imaging}. A C-scan pixel is the reflected amplitude extracted at one chosen depth and therefore carries a depth hypothesis: if the selected slice does not coincide with the void, the contrast weakens even when the void is present, and the appropriate slice differs between specimens and between bandwidths. A pixel of $\mathbf{P}$ is a void posterior computed from the whole sub-band at that position, so no depth is selected at any point in the procedure. Explicit high-resolution range-profile reconstruction, deconvolution, and source-wavelet estimation are likewise not required.

Two properties of~\eqref{eq:probmap} are relevant to practical inspection. The mapping is point-wise, so it inherits whatever spatial sampling the scan provides and can be applied to a different scanning pattern or to a partially scanned facade without modification. It also consumes only $B_{\mathrm{sub}}$ of bandwidth at each position, which, as argued in Section~\ref{sec:intro}, is the single term that sets front-end cost, acquisition time, and regulatory feasibility in the field; the framework therefore trades hardware bandwidth for computation. Whether a decision of useful quality can in fact be made from so narrow a band is an empirical question, and Section~\ref{sec:results} answers it by evaluating the framework from 16\,GHz down to 0.5\,GHz under the mixed-sample protocol and from 2\,GHz down to 0.5\,GHz under the leave-one-specimen-out protocol.

\section{Experimental Results and Discussion}\label{sec:results}

\subsection{Experimental Protocols}\label{ssec:protocols}

All sub-band settings are centered on 24\,GHz. This center frequency was chosen
because the 24\,GHz band is mature both in regulatory and in commercial terms:
short-range radar operation is permitted around it in most jurisdictions, and
compact, low-cost front ends operating there are widely available. The exact
allocation in the vicinity of 24\,GHz nevertheless differs between
jurisdictions, so an inspection system intended for deployment across regions
cannot assume that a wide contiguous allocation will be available everywhere.
The settings therefore hold the center frequency at 24\,GHz and reduce the
occupied bandwidth as far as possible, the narrowest spanning
23.75--24.25\,GHz. Holding the center fixed has the further benefit of
isolating the effect of bandwidth from that of the operating frequency, since
the settings then differ only in width. The narrowband imaging example of
Section~\ref{sec:imaging} uses 24--26\,GHz, which has the same 2\,GHz width as
one of these settings and lies in the same part of the spectrum, so the two
analyses are comparable in bandwidth. The same preprocessing procedure and
narrowband RF representation described in Section~\ref{sec:method} were used
for all bandwidth settings.

Two evaluation protocols were considered. In the first protocol, referred to as the \emph{mixed-sample split}, data from all wall specimens were randomly divided into training, validation, and test sets with a ratio of 6:2:2. This setting evaluates whether the model can distinguish void and non-void RF responses when the training and testing data follow the same specimen distribution. In the second protocol, referred to as the \emph{leave-one-specimen-out split}, 11 specimens were used for training and validation while the remaining specimen was used only for testing, with the training and validation data split at a ratio of 8:2. The procedure is repeated with each of the twelve specimens held out in turn, so every figure reported below for this protocol is aggregated over twelve folds. This protocol provides a stricter evaluation of generalization because the model must detect voids in a wall specimen that is not included during training.

Each specimen is scanned on a $60\times60$ grid, giving 3{,}600 positions, of which the boundary-region exclusion described below removes 480. The remaining 3{,}120 positions comprise 292 void and 2{,}828 normal positions, a normal-to-void ratio of approximately $9.7\!:\!1$, so the twelve specimens together provide 37{,}440 labeled positions of which 3{,}504 are void. At the narrowest 0.5\,GHz setting each position carries $N_f=28$ of the 2{,}048 measured frequency points. Every value reported for the mixed-sample split is the mean and standard deviation over three independent training runs with different random seeds, whereas every value reported for the leave-one-specimen-out split is the mean and standard deviation over the twelve folds. All percentages are reported to two decimal places.

The training labels were generated based on the known positions of the artificial voids specified by the specimen design. Owing to the fine spatial sampling of the measurement system, regions inside the manually defined void boundaries were labeled as void samples, whereas regions outside the void boundaries were labeled as normal samples. Since the RF response near a void boundary can be affected by mixed contributions from the void and surrounding normal regions, an approximately 10-mm-wide boundary region around each void edge was excluded from training, as shown in Fig.~\ref{fig:pipeline}. This exclusion reduces ambiguous labels and improves the reliability of supervised learning.

Because the number of normal samples is much larger than that of void samples, the F1-score was used as the main evaluation metric. Accuracy, precision, and recall were also considered to provide a more complete performance evaluation.

\subsection{Mixed-Sample Detection Performance}\label{ssec:mixed}

The mixed-sample split was first used to evaluate the basic discriminative capability of the proposed narrowband RF learning method. In this setting, scan positions from all specimens were included in the training, validation, and test sets, allowing the model to learn void-related RF patterns under the same overall specimen distribution.

\begin{figure}
  \centering
  \includegraphics[width=\linewidth]{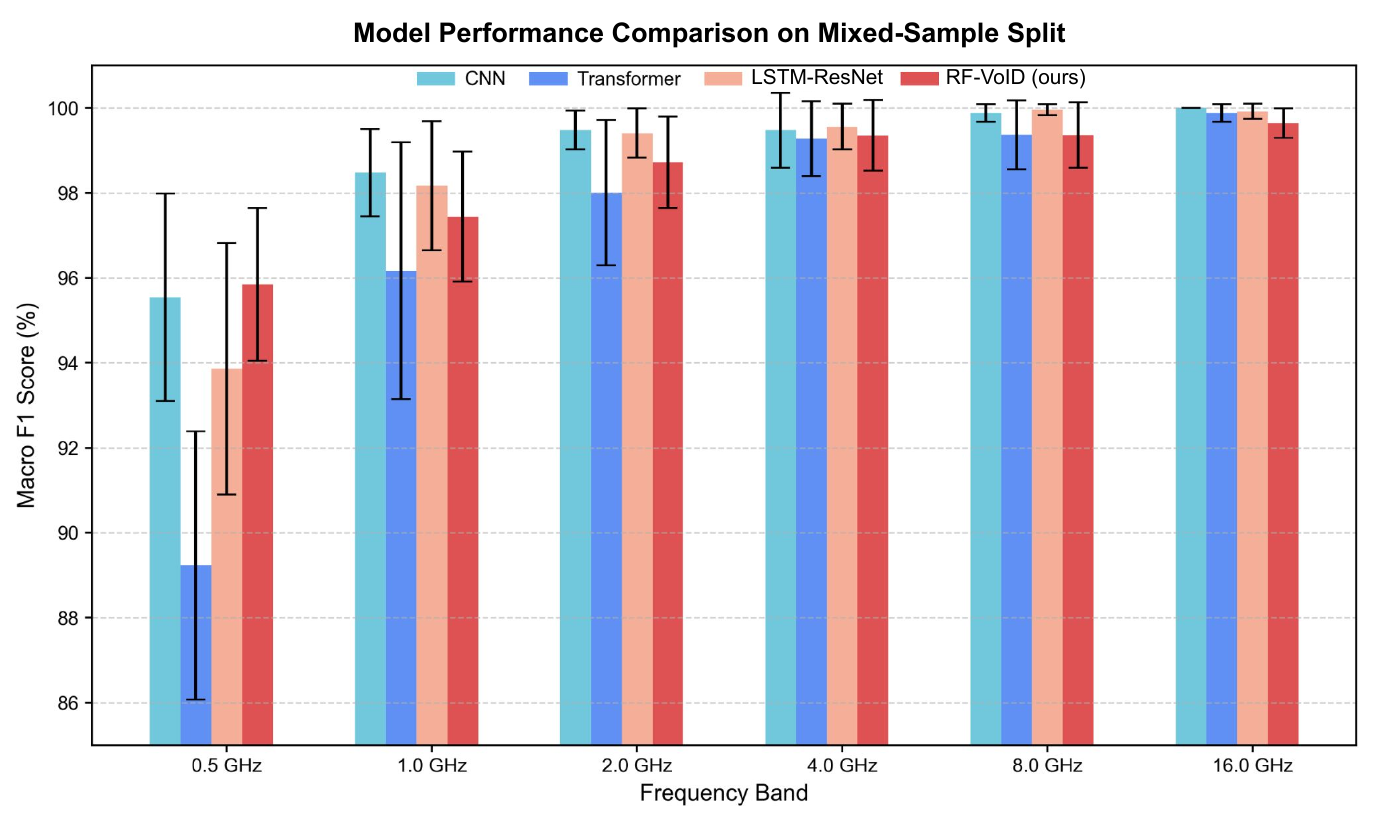}
  \hspace{-3mm}
  \caption{Model performance under the mixed-sample split using different models.}
  \label{fig:mixed_bar}
\end{figure}

Fig.~\ref{fig:mixed_bar} compares the quantitative performance of different models under the mixed-sample split. The proposed method achieves stable detection performance across different bandwidth settings, indicating that the narrowband complex RF response contains useful void-sensitive information even when conventional range-domain visualization becomes degraded.

As summarized in Table~\ref{tbl:mixed}, the quantitative results clarify how detection performance changes as the available bandwidth is reduced. A gradual performance decrease is expected at narrower bandwidths because the range-domain response becomes broader and neighboring echoes increasingly overlap. However, the model retains a 97.44\% macro F1-score and 95.36\% void-class F1 at 1\,GHz, which supports the central hypothesis of this work: thin voids can still perturb the frequency-domain response even when they are not clearly separable in conventional B-scan or C-scan images.

\begin{table}[tbp]
  \centering
\scalebox{0.84}{
\begin{tabular}{l|l|cccc}
\toprule
BW.     & Freq. (GHz) & Acc.                                                       & F1                                                         & F1(N)                                               & F1(V)                                                 \\ \midrule
16\,GHz  & 16.0--32.0   & 99.88 & 99.64 & 99.93 & 99.35 \\
         &              & $\pm$0.12 & $\pm$0.35 & $\pm$0.07 & $\pm$0.63 \\
8\,GHz   & 20.0--28.0   & 99.79 & 99.36 & 99.88 & 98.84 \\
         &              & $\pm$0.26 & $\pm$0.77 & $\pm$0.14 & $\pm$1.40 \\
4\,GHz   & 22.0--26.0   & 99.79 & 99.36 & 99.88 & 98.83 \\
         &              & $\pm$0.27 & $\pm$0.83 & $\pm$0.15 & $\pm$1.50 \\
2\,GHz   & 23.0--25.0   & 99.57 & 98.72 & 99.76 & 97.68 \\
         &              & $\pm$0.36 & $\pm$1.08 & $\pm$0.20 & $\pm$1.96 \\
1\,GHz   & 23.5--24.5   & 99.15 & 97.44 & 99.53 & 95.36 \\
         &              & $\pm$0.50 & $\pm$1.53 & $\pm$0.28 & $\pm$2.78 \\
0.5\,GHz & 23.75--24.25 & 98.61 & 95.84 & 99.24 & 92.45 \\
         &              & $\pm$0.61 & $\pm$1.80 & $\pm$0.34 & $\pm$3.27 \\ \bottomrule
\end{tabular}}
\caption{Mixed-sample split detection performance under different bandwidths. \textit{Acc.} refers to accuracy, macro F1 is applied as the \textit{F1} score here, \textit{F1(N)} and \textit{F1(V)} stand for F1 scores of the normal and void classes, respectively.}
\label{tbl:mixed}
\end{table}

\begin{figure}
  \centering
  \includegraphics[width=\linewidth]{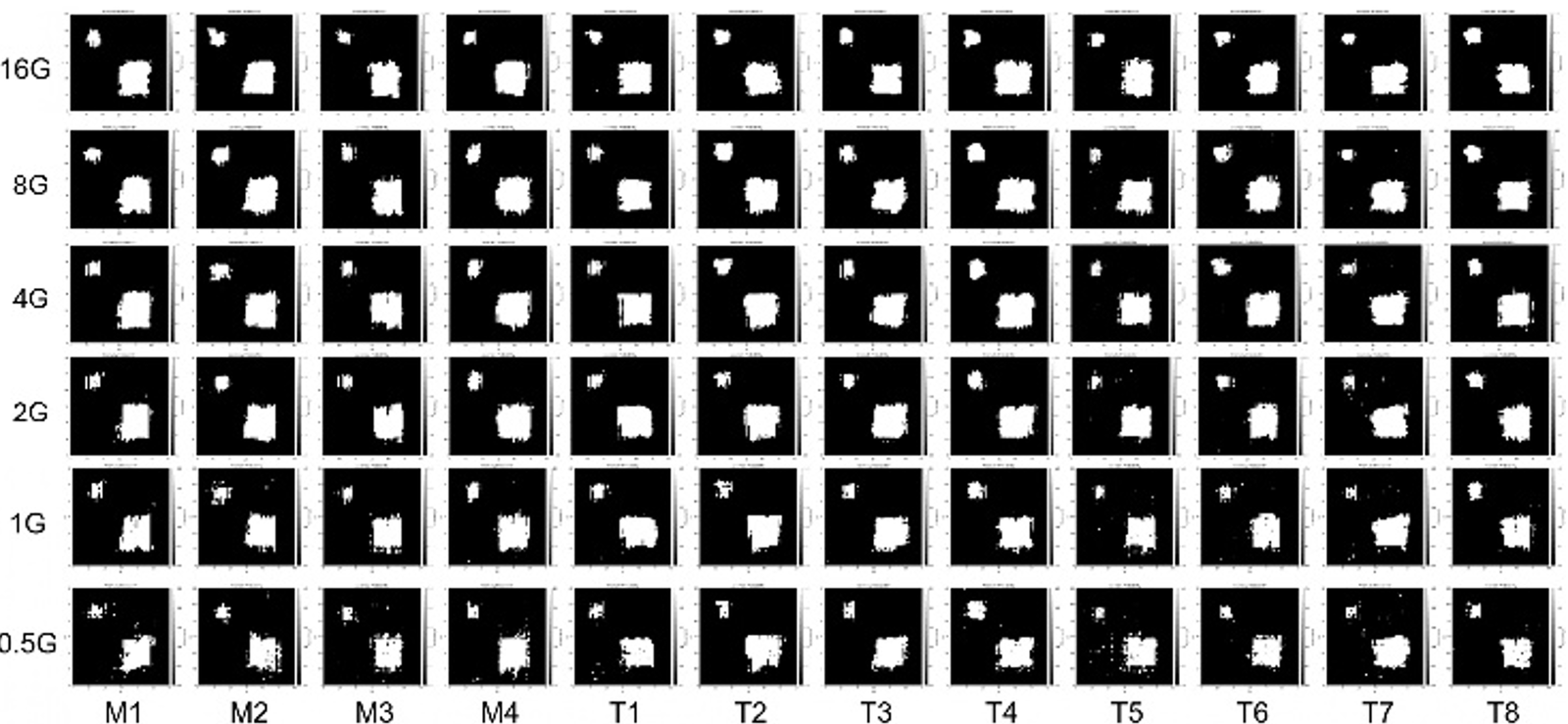}
  \caption{Defect probability maps of the proposed method across different bandwidths under the mixed-sample split.}
  \label{fig:mixed_maps}
\end{figure}

Fig.~\ref{fig:mixed_maps} shows the defect probability maps obtained under different bandwidth settings. The rows correspond to different bandwidths, while the columns correspond to different exterior-wall specimens, including the M-series and T-series samples. At larger bandwidths, such as 16\,GHz and 8\,GHz, the predicted void regions are generally cleaner and more spatially consistent. As the bandwidth decreases, the prediction maps become noisier and the void boundaries become less stable. Nevertheless, the main void locations can still be identified in many cases, even under the 0.5\,GHz bandwidth.

This result is important from a practical inspection perspective. Conventional radar imaging typically requires a wide bandwidth to obtain sufficient range resolution for stable visualization of thin voids. In contrast, the proposed AI-based method can still generate useful void probability maps from narrowband complex RF responses, suggesting the possibility of bandwidth-efficient exterior-wall inspection.

\begin{figure*}[t]
  \centering
  \includegraphics[width=0.8\textwidth]{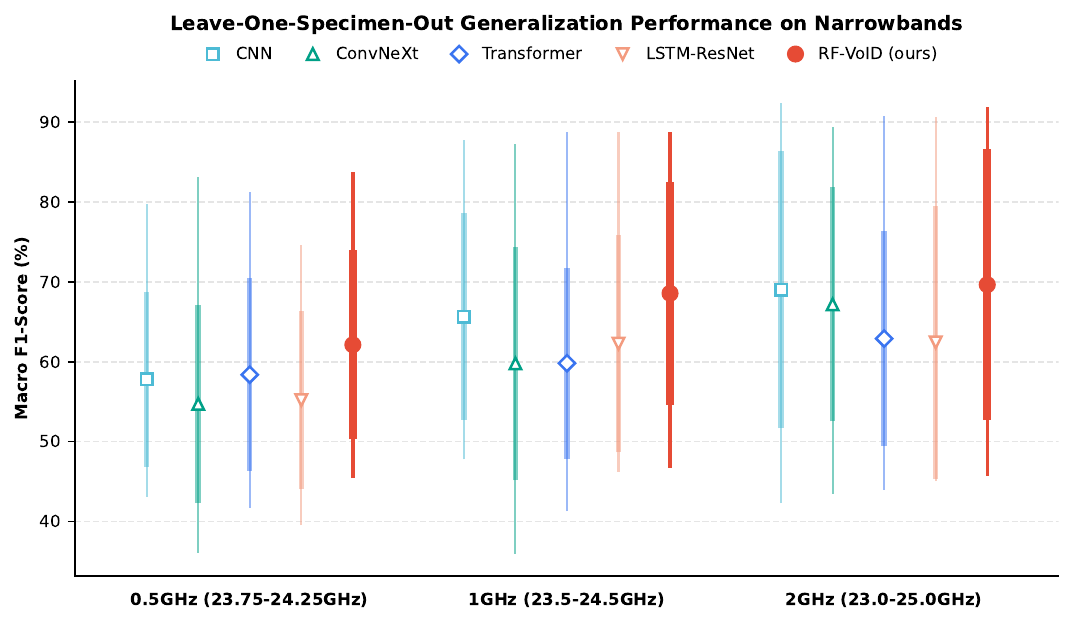}
  \caption{Leave-one-specimen-out generalization performance of different models under 0.5 GHz, 1 GHz and 2 GHz bandwidths. The markers indicate the mean macro F1-score over unseen test specimens; the thick bars span one standard deviation about the mean and the thin bars span the full min--max range across the twelve leave-one-specimen-out folds.}
  \label{fig:sample_bar}
\end{figure*}

\subsection{Leave-One-Specimen-Out Generalization Performance}\label{ssec:sample}

To further evaluate generalization, the leave-one-specimen-out split was used. In this experiment, 11 specimens were used for training and validation, and one unseen specimen was used only for testing. This leave-one-specimen-out protocol is more challenging than the mixed-sample split because the test specimen is completely excluded from training. Therefore, the model must detect voids under specimen-dependent variations in material properties, multilayer structure, void condition, and measurement conditions.

As summarized in Table~\ref{tbl:sample}, a performance decrease compared with the mixed-sample split is expected, since the domain gap between training and testing specimens must now be bridged by the learned features alone. The results under this protocol should therefore be read as the more realistic assessment of whether those features transfer to new wall specimens.

Fig.~\ref{fig:sample_bar} compares the leave-one-specimen-out generalization performance of different models under the 0.5\,GHz, 1\,GHz, and 2\,GHz bandwidth settings. The macro F1-score was used because the number of normal scan points is much larger than that of void points. Compared with the baseline CNN~\citep{726791}, ConvNeXt~\citep{Liu2022}, Transformer~\citep{Vaswani2017}, and LSTM-ResNet model~\citep{choi2018short}, the proposed RF-VoID achieves the highest mean macro F1-score under all three bandwidth settings. This result suggests that the proposed narrowband RF representation and model design improve the ability to detect void-related responses in unseen specimens.

Extending the evaluation to 0.5\,GHz shows where bandwidth begins to matter under this protocol. Between 2\,GHz and 1\,GHz the mean macro F1-score shows no detectable change (exact two-sided Wilcoxon signed-rank test over the twelve paired folds, $W=28$, $p=0.42$), whereas narrowing further to 0.5\,GHz lowers it by 6.45 points on average, with ten of the twelve held-out specimens scoring lower than at 1\,GHz ($W=3$, $p=0.002$). The per-specimen scores remain strongly correlated across bandwidths (Pearson $r=0.93$ between 0.5\,GHz and 1\,GHz), so the ordering of easy and difficult specimens is largely preserved and the loss at 0.5\,GHz is broadly systematic rather than confined to a few walls, although its magnitude varies substantially between specimens, from $+1.9$ to $-15.9$ points. The same pattern appears in the mixed-sample results of Table~\ref{tbl:mixed}, where the largest single step in both macro and void-class F1 also occurs between 1\,GHz and 0.5\,GHz; under domain shift that step is simply larger.

\begin{table}[h]
\centering
\scalebox{0.66}{
\begin{tabular}{l|cc|cc|cc}
\toprule
 & \multicolumn{2}{c|}{0.5 GHz (23.75--24.25)} & \multicolumn{2}{c|}{1 GHz (23.5--24.5)} & \multicolumn{2}{c}{2 GHz (23.0--25.0)} \\ \midrule
Spec.  & Acc. & F1 & Acc. & F1 & Acc. & F1 \\ \midrule
M1    & 83.17 & 45.41 & 88.11 & 49.63 & 80.96 & 45.72 \\
M2    & 77.98 & 48.82 & 83.27 & 46.91 & 74.87 & 50.99 \\
M3    & 88.56 & 55.32 & 91.19 & 65.73 & 90.45 & 54.66 \\
M4    & 85.29 & 64.44 & 86.44 & 68.77 & 89.68 & 72.05 \\
T1    & 85.48 & 61.98 & 93.21 & 77.92 & 96.28 & 88.20 \\
T2    & 93.27 & 80.87 & 96.28 & 87.44 & 97.50 & 91.87 \\
T3    & 93.11 & 72.13 & 93.04 & 74.69 & 94.29 & 77.77 \\
T4    & 94.01 & 83.74 & 95.83 & 88.84 & 96.60 & 90.62 \\
T5    & 88.85 & 55.38 & 90.42 & 60.34 & 89.87 & 52.17 \\
T6    & 90.29 & 48.73 & 88.46 & 48.02 & 89.84 & 50.00 \\
T7    & 89.87 & 67.86 & 92.47 & 79.51 & 92.21 & 78.27 \\
T8    & 90.48 & 60.79 & 93.37 & 75.06 & 94.94 & 83.48 \\ \midrule
\begin{tabular}[c]{@{}l@{}}Mean \\ $\pm$Std.\end{tabular} & \begin{tabular}[c]{@{}c@{}}88.36\\ $\pm$4.49\end{tabular} & \begin{tabular}[c]{@{}c@{}}62.12\\ $\pm$11.85\end{tabular} & \begin{tabular}[c]{@{}c@{}}91.01\\ $\pm$3.69\end{tabular} & \begin{tabular}[c]{@{}c@{}}68.57\\ $\pm$14.08\end{tabular} & \begin{tabular}[c]{@{}c@{}}90.62\\ $\pm$6.41\end{tabular} & \begin{tabular}[c]{@{}c@{}}69.65\\ $\pm$16.96\end{tabular} \\ \bottomrule
\end{tabular}}
\caption{Leave-one-specimen-out generalization performance of RF-VoID under narrow bandwidths. For each row, the indicated specimen is only used for testing and the remaining specimens are used for training.}
\label{tbl:sample}
\end{table}

\begin{figure}[h]
  \centering
  \includegraphics[width=\linewidth]{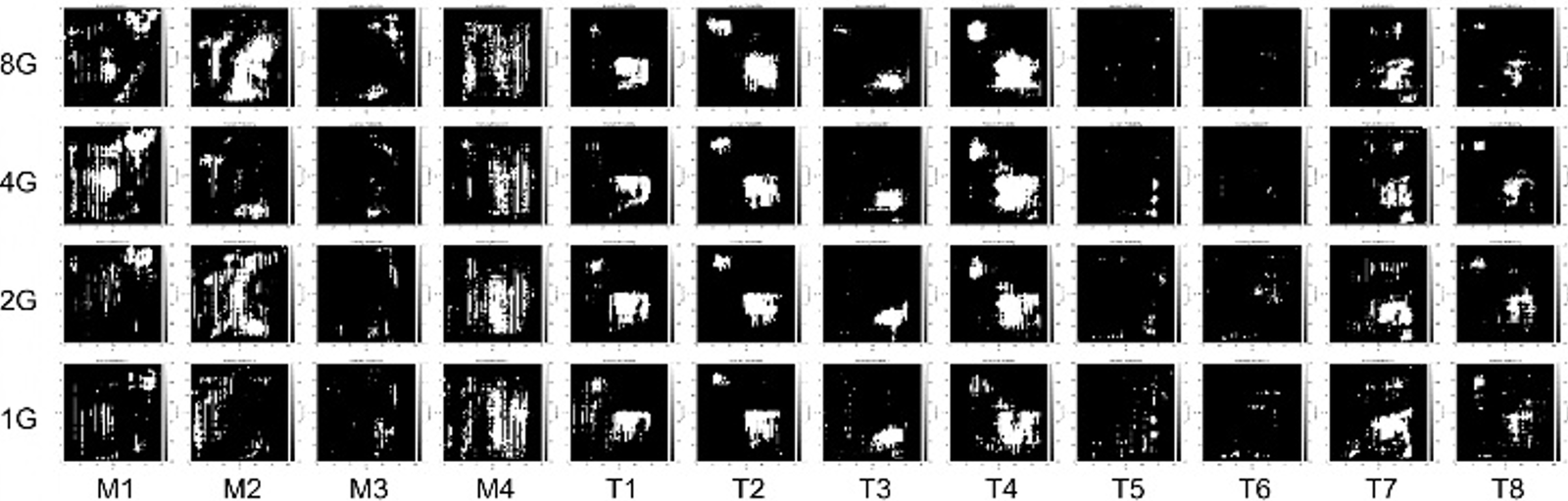}
  \caption{Defect probability maps obtained under the leave-one-specimen-out split. The results illustrate the leave-one-specimen-out generalization capability of the proposed model under different bandwidth conditions.}
  \label{fig:sample_maps}
\end{figure}

The error bars in Fig.~\ref{fig:sample_bar} indicate that the leave-one-specimen-out performance varies considerably across unseen specimens. This variation is expected because each specimen differs in multilayer structure, material response, void position, and local scattering condition. The spread of macro F1 across folds is widest at 2\,GHz and narrowest at 0.5\,GHz. At every setting this spread exceeds the mean difference between settings, so the bandwidth effect identified above is visible only in the paired comparison, in which each specimen serves as its own control, and not in the overlap of the error bars; the dominant source of variance under this protocol remains which specimen is held out.

Fig.~\ref{fig:sample_maps} presents the corresponding defect probability maps obtained in the leave-one-specimen-out setting. Compared with the mixed-sample results, the probability maps are generally less stable, especially at narrower bandwidths. Some specimens show clear localization of the void regions, whereas others contain false detections or missed regions. This behavior indicates that the proposed model can extract transferable void-related features from narrowband RF responses for several unseen specimens, but it also reveals that generalization remains dependent on the diversity of training specimens.

The comparison between the mixed-sample and leave-one-specimen-out splits highlights an important practical limitation. Although the proposed method enables narrowband void detection, its robustness is affected by specimen-dependent variations. Therefore, future work should include additional wall specimens with different tile materials, mortar layers, concrete substrates, void depths, measurement distances, and surface conditions. Such data diversity will be important for improving the reliability of AI-based millimeter-wave inspection in practical exterior-wall applications.

\subsection{Ablation Study}\label{ssec:ablation}

To further clarify the contribution of the proposed input representation, an ablation study is conducted under the mixed-sample split. The purpose of this experiment is to verify the necessity of the physics-informed multi-channel input. All input variants were trained and evaluated under the same 0.5-GHz mixed-sample setting, and every reported value is the mean and standard deviation over the same three independent training runs used in Section~\ref{ssec:mixed}.

The input variants evaluated include in-phase only, quadrature only, raw I/Q (real and imaginary), and the proposed full complex RF representation (real, imaginary, magnitude, and phase).

\begin{table}[tbp]
\centering
\scalebox{0.86}{
\begin{tabular}{l|cccc}
\toprule
\multirow{2}{*}{Input Variants} & \multicolumn{4}{c}{0.5 GHz (23.75--24.25)}         \\ \cline{2-5} 
                                & Acc.       & F1         & F1(N)      & F1(V)      \\ \midrule
In-phase Input                  & \begin{tabular}[c]{@{}c@{}}98.23\\ $\pm$0.75\end{tabular} & \begin{tabular}[c]{@{}c@{}}94.66\\ $\pm$2.27\end{tabular} & \begin{tabular}[c]{@{}c@{}}99.02\\ $\pm$0.41\end{tabular} & \begin{tabular}[c]{@{}c@{}}90.30\\ $\pm$4.14\end{tabular} \\ \midrule
Quadrature Input                & \begin{tabular}[c]{@{}c@{}}98.12\\ $\pm$1.04\end{tabular} & \begin{tabular}[c]{@{}c@{}}94.34\\ $\pm$3.16\end{tabular} & \begin{tabular}[c]{@{}c@{}}98.96\\ $\pm$0.57\end{tabular} & \begin{tabular}[c]{@{}c@{}}89.71\\ $\pm$5.76\end{tabular} \\ \midrule
I/Q Input                       & \begin{tabular}[c]{@{}c@{}}98.36\\ $\pm$0.69\end{tabular} & \begin{tabular}[c]{@{}c@{}}95.00\\ $\pm$2.17\end{tabular} & \begin{tabular}[c]{@{}c@{}}99.10\\ $\pm$0.38\end{tabular} & \begin{tabular}[c]{@{}c@{}}90.91\\ $\pm$3.96\end{tabular} \\ \midrule
Complex RF Input                & \begin{tabular}[c]{@{}c@{}}98.61\\ $\pm$0.61\end{tabular} & \begin{tabular}[c]{@{}c@{}}95.84\\ $\pm$1.80\end{tabular} & \begin{tabular}[c]{@{}c@{}}99.24\\ $\pm$0.34\end{tabular} & \begin{tabular}[c]{@{}c@{}}92.45\\ $\pm$3.27\end{tabular} \\ \bottomrule
\end{tabular}}
\caption{Ablation study of input variants on 0.5 GHz narrowband.}
\label{tbl:ablation}
\end{table}

As shown in Table~\ref{tbl:ablation}, relying on a single-channel input (either in-phase or quadrature) or even the raw dual-channel I/Q input yields suboptimal performance under the 0.5-GHz bandwidth setting. In contrast, the proposed full complex RF input consistently outperforms these simplified representations. This demonstrates that thin air-void detection under narrowband conditions critically benefits from explicitly combining amplitude, phase, and raw complex-valued frequency-domain information.

\section{Conclusion}\label{sec:conclusion}

This paper addressed hidden-void detection behind exterior ceramic tiles under
a constrained sensing bandwidth. Conventional A-scan, B-scan, and C-scan
interpretation was analyzed first, and that analysis established the bound
which motivates the rest of the work: because the depth resolution of a
stepped-frequency measurement is inversely proportional to bandwidth, a
0.5--1.0\,mm void is already unresolved at the full 4--40\,GHz sweep, and
narrowing the band only widens the overlap between the void response and the
surface and interface echoes.

RF-VoID was then introduced to move the decision out of the image domain.
The sub-band complex response is kept in its measured frequency order,
amplitude and phase are computed alongside the in-phase and quadrature
channels, and a dual-branch encoder reads the resulting sequence along the
frequency axis, with relative position encoding and a distance-dependent
locality bias in the attention branch and an inspection-oriented objective
that accounts for class imbalance and the asymmetric cost of a missed void.
Under the mixed-sample protocol the framework reaches 98.61\% accuracy and a
95.84\% overall F1-score at 0.5\,GHz, a seventy-two-fold reduction in occupied
bandwidth relative to the full sweep, with no range-profile reconstruction, no
deconvolution, and no depth-slice selection at any point.

These results support the claim made at the outset: for this task the
bandwidth an inspection requires can be traded for computation, and it is
bandwidth rather than detection performance that sets the cost, the
acquisition time, and the regulatory footprint of a deployed system. The trade
is not yet complete. Under the leave-one-specimen-out protocol the performance
falls well below the mixed-sample figures and varies considerably between
held-out specimens, which shows that the features learned here are only partly
transferable across walls differing in layer thickness, material, and surface
condition. Bandwidth also re-enters under this protocol: performance is
indistinguishable at 2\,GHz and 1\,GHz but drops significantly at 0.5\,GHz,
so the narrowest setting that suffices when the wall is represented in
training is not yet sufficient for a wall that is not. Closing that gap is the main obstacle to field use, and it is a
data problem before it is a modeling one: the next step is to extend the
specimen set across tile materials, mortar formulations, substrate types, void
depths, stand-off distances, and outdoor surface conditions.

\section*{Acknowledgements}

This work was supported by JST BOOST, Japan [grant number JPMJBY25A1]. The
authors gratefully acknowledge TOF Corporation, Japan, for providing the
samples and equipment used in this study.

\section*{Data availability}

Data will be made available on request.

\bibliographystyle{assets/plainnat}
\bibliography{cas-refs}

@article{Soeta2016,
  author  = {T. Soeta and T. Mikami},
  title   = {Basic study into a diagnostic system for exterior tile debonding},
  journal = {Journal of Structural and Construction Engineering (Transactions of AIJ)},
  volume  = {81},
  number  = {729},
  pages   = {1779--1787},
  year    = {2016},
  doi     = {10.3130/aijs.81.1779}
}

@article{Ito2025,
  author  = {A. Ito and M. Koike and M. Saito and K. Hibino},
  title   = {Hammering test for tile wall using deep learning},
  journal = {Applied Sciences},
  volume  = {15},
  number  = {3},
  pages   = {1500},
  year    = {2025},
  doi     = {10.3390/app15031500}
}

@article{Li2000,
  author  = {Z. Li and W. Yao and S. Lee and C. Lee and Z. Yang},
  title   = {Application of infrared thermography technique in building finish
             evaluation},
  journal = {Journal of Nondestructive Evaluation},
  volume  = {19},
  number  = {1},
  pages   = {11--19},
  year    = {2000},
  doi     = {10.1023/A:1006612023656}
}

@article{Lourenco2017,
  author  = {T. Lourenco and L. Matias and P. Faria},
  title   = {Anomalies detection in adhesive wall tiling systems by infrared
             thermography},
  journal = {Construction and Building Materials},
  volume  = {148},
  pages   = {419--428},
  year    = {2017},
  doi     = {10.1016/j.conbuildmat.2017.05.052}
}

@article{Kharkovsky2007,
  author  = {S. Kharkovsky and R. Zoughi},
  title   = {Microwave and millimeter wave nondestructive testing and evaluation:
             Overview and recent advances},
  journal = {IEEE Instrumentation \& Measurement Magazine},
  volume  = {10},
  number  = {2},
  pages   = {26--38},
  year    = {2007},
  doi     = {10.1109/MIM.2007.364985}
}

@article{Brinker2020,
  author  = {K. Brinker and M. Dvorsky and M. T. {Al Qaseer} and R. Zoughi},
  title   = {Review of advances in microwave and millimetre-wave {NDT\&E}:
             Principles and applications},
  journal = {Philosophical Transactions of the Royal Society A},
  volume  = {378},
  number  = {2182},
  pages   = {20190585},
  year    = {2020},
  doi     = {10.1098/rsta.2019.0585}
}

@article{Yi2018,
  author  = {L. Yi and L. Zou and K. Takahashi and M. Sato},
  title   = {High-resolution velocity analysis method using the l1-norm
             regularized least-squares method for pavement inspection},
  journal = {IEEE Journal of Selected Topics in Applied Earth Observations and
             Remote Sensing},
  volume  = {11},
  number  = {3},
  pages   = {1005--1015},
  year    = {2018},
  doi     = {10.1109/JSTARS.2018.2797018}
}

@article{Chen2020,
  author  = {H. Chen and Z. Long and L. Niu and Z. Yang and J. Liu and K. Wang},
  title   = {Millimeter-wave {SFCW} {SAR} imaging system based on in-phase
             signal measurement with simplified transceiver},
  journal = {Optics Express},
  volume  = {28},
  number  = {2},
  pages   = {1526--1538},
  year    = {2020},
  doi     = {10.1364/OE.380266}
}

@inproceedings{Koyabu2023,
  author    = {Y. Koyabu and H. Tokunaga and Y. Wang and Y. Li and T. Ohtsuka
               and Y. Tagami and T. Nagatsuma},
  title     = {Ultrawide-band millimeter-wave photonic radar mounted on drone},
  booktitle = {Proc. 2023 Asia-Pacific Microwave Conference (APMC)},
  address   = {Taipei, Taiwan},
  pages     = {500--502},
  year      = {2023}
}

@inproceedings{Alsalem2020,
  author    = {H. Alsalem and T. Tanaka and T. Honda and S. Doi and S. Uchida},
  title     = {Measuring adhesion strength of wall tile to concrete by
               non-contact inspection using electromagnetic waves},
  booktitle = {Proc. 37th International Symposium on Automation and Robotics
               in Construction (ISARC)},
  address   = {Kitakyushu, Japan},
  pages     = {633--638},
  year      = {2020}
}

@inproceedings{Shoda2025,
  author    = {S. Shoda and R. Ma and L. Yi},
  title     = {Exterior wall inspection using millimeter-wave radar: A
               bandwidth-efficient approach with deconvolution},
  booktitle = {Proc. 2025 9th Asia-Pacific Conference on Synthetic Aperture
               Radar (APSAR)},
  address   = {Matsue, Japan},
  year      = {2025},
  doi       = {10.23919/APSAR64635.2025.11392578}
}

@article{Wang2019,
  author  = {P. Wang and Z. Li and P. Liu and Y. Pei},
  title   = {Super resolution in depth for microwave imaging},
  journal = {Applied Physics Letters},
  volume  = {115},
  number  = {4},
  pages   = {044101},
  year    = {2019},
  doi     = {10.1063/1.5098302}
}

@article{Candes2006,
  author  = {E. J. Candes and J. Romberg and T. Tao},
  title   = {Robust uncertainty principles: Exact signal reconstruction from
             highly incomplete frequency information},
  journal = {IEEE Transactions on Information Theory},
  volume  = {52},
  number  = {2},
  pages   = {489--509},
  year    = {2006},
  doi     = {10.1109/TIT.2005.862083}
}

@article{Donoho2006,
  author  = {D. L. Donoho},
  title   = {Compressed sensing},
  journal = {IEEE Transactions on Information Theory},
  volume  = {52},
  number  = {4},
  pages   = {1289--1306},
  year    = {2006},
  doi     = {10.1109/TIT.2006.871582}
}

@article{Potter2010,
  author  = {L. C. Potter and E. Ertin and J. T. Parker and M. Cetin},
  title   = {Sparsity and compressed sensing in radar imaging},
  journal = {Proceedings of the IEEE},
  volume  = {98},
  number  = {6},
  pages   = {1006--1020},
  year    = {2010},
  doi     = {10.1109/JPROC.2009.2037526}
}

@misc{Gao2018,
  author = {J. Gao and B. Deng and Y. Qin and H. Wang and X. Li},
  title  = {Fast super-resolution {3D} {SAR} imaging using an unfolded deep
            network},
  year   = {2018},
  note   = {arXiv:1808.08658},
  doi    = {10.48550/arXiv.1808.08658}
}

@article{Zhu2021,
  author  = {X. X. Zhu and S. Montazeri and M. Ali and Y. Hua and Y. Wang and
             L. Mou and Y. Shi and F. Xu and R. Bamler},
  title   = {Deep learning meets {SAR}: Concepts, models, pitfalls, and
             perspectives},
  journal = {IEEE Geoscience and Remote Sensing Magazine},
  volume  = {9},
  number  = {4},
  pages   = {143--172},
  year    = {2021},
  doi     = {10.1109/MGRS.2020.3046356}
}

@article{Moreira2013,
  author  = {A. Moreira and P. Prats-Iraola and M. Younis and G. Krieger and
             I. Hajnsek and K. P. Papathanassiou},
  title   = {A tutorial on synthetic aperture radar},
  journal = {IEEE Geoscience and Remote Sensing Magazine},
  volume  = {1},
  number  = {1},
  pages   = {6--43},
  year    = {2013},
  doi     = {10.1109/MGRS.2013.2248301}
}

@article{Sheen2001,
  author  = {D. M. Sheen and D. L. McMakin and T. E. Hall},
  title   = {Three-dimensional millimeter-wave imaging for concealed weapon
             detection},
  journal = {IEEE Transactions on Microwave Theory and Techniques},
  volume  = {49},
  number  = {9},
  pages   = {1581--1592},
  year    = {2001},
  doi     = {10.1109/22.942570}
}

@article{Robert1998,
  author  = {A. Robert},
  title   = {Dielectric permittivity of concrete between 50~{MHz} and 1~{GHz}
             and {GPR} measurements for building materials evaluation},
  journal = {Journal of Applied Geophysics},
  volume  = {40},
  number  = {1--3},
  pages   = {89--94},
  year    = {1998},
  doi     = {10.1016/S0926-9851(98)00009-3}
}

@article{Maierhofer2003,
  author  = {C. Maierhofer},
  title   = {Nondestructive evaluation of concrete infrastructure with ground
             penetrating radar},
  journal = {Journal of Materials in Civil Engineering},
  volume  = {15},
  number  = {3},
  pages   = {287--297},
  year    = {2003},
  doi     = {10.1061/(ASCE)0899-1561(2003)15:3(287)}
}

@book{Pozar2012,
  author    = {D. M. Pozar},
  title     = {Microwave Engineering},
  edition   = {4th},
  publisher = {Wiley},
  address   = {Hoboken, NJ, USA},
  year      = {2012}
}

@article{Zoughi1990,
  author  = {R. Zoughi and S. Bakhtiari},
  title   = {Microwave nondestructive detection and evaluation of disbonding
             and delamination in layered-dielectric slabs},
  journal = {IEEE Transactions on Instrumentation and Measurement},
  volume  = {39},
  number  = {6},
  pages   = {1059--1063},
  year    = {1990},
  doi     = {10.1109/19.65826}
}

@article{Wu2020,
  author  = {C. Wu and F. Zhang and B. Wang and K. J. R. Liu},
  title   = {{mSense}: Towards mobile material sensing with a single
             millimeter-wave radio},
  journal = {Proceedings of the ACM on Interactive, Mobile, Wearable and
             Ubiquitous Technologies},
  volume  = {4},
  number  = {3},
  pages   = {106},
  year    = {2020},
  doi     = {10.1145/3411822}
}

@inproceedings{Chen2026,
  author    = {X. Chen and Q. Li and R. Ma and J. Bai and Y. Li and J. Yang},
  title     = {{RF-MatID}: Dataset and benchmark for radio frequency material
               identification},
  booktitle = {Proc. Int. Conf. Learn. Represent. (ICLR)},
  year      = {2026}
}

@book{Richards2014,
  author    = {M. A. Richards},
  title     = {Fundamentals of Radar Signal Processing},
  edition   = {2nd},
  publisher = {McGraw-Hill},
  address   = {New York, NY, USA},
  year      = {2014}
}

@article{Su2024,
  author  = {J. Su and Y. Lu and S. Pan and A. Murtadha and B. Wen and Y. Liu},
  title   = {{RoFormer}: Enhanced transformer with rotary position embedding},
  journal = {Neurocomputing},
  volume  = {568},
  pages   = {127063},
  year    = {2024}
}

@inproceedings{Press2022,
  author    = {O. Press and N. A. Smith and M. Lewis},
  title     = {Train short, test long: Attention with linear biases enables
               input length extrapolation},
  booktitle = {Proc. Int. Conf. Learn. Represent. (ICLR)},
  year      = {2022}
}

@inproceedings{Vaswani2017,
  author    = {Vaswani, Ashish and Shazeer, Noam and Parmar, Niki and
               Uszkoreit, Jakob and Jones, Llion and Gomez, Aidan N. and
               Kaiser, {\L}ukasz and Polosukhin, Illia},
  title     = {Attention is all you need},
  booktitle = {Advances in Neural Information Processing Systems (NeurIPS)},
  volume    = {30},
  pages     = {5998--6008},
  year      = {2017}
}

@inproceedings{Liu2022,
  author    = {Liu, Zhuang and Mao, Hanzi and Wu, Chao-Yuan and
               Feichtenhofer, Christoph and Darrell, Trevor and Xie, Saining},
  title     = {A {ConvNet} for the 2020s},
  booktitle = {Proceedings of the IEEE/CVF Conference on Computer Vision and
               Pattern Recognition (CVPR)},
  pages     = {11976--11986},
  year      = {2022},
  doi       = {10.1109/CVPR52688.2022.01167}
}

@article{Lin2017,
  author  = {Lin, Tsung-Yi and Goyal, Priya and Girshick, Ross and
             He, Kaiming and Doll{\'a}r, Piotr},
  title   = {Focal loss for dense object detection},
  journal = {IEEE Transactions on Pattern Analysis and Machine Intelligence},
  volume  = {42},
  number  = {2},
  pages   = {318--327},
  year    = {2020},
  doi     = {10.1109/TPAMI.2018.2858826}
}

@article{Liu2020,
  author  = {Liu, Hai and Lin, Chunxu and Cui, Jie and Fan, Lisheng and
             Xie, Xiongyao and Spencer, Billie F.},
  title   = {Detection and localization of rebar in concrete by deep learning
             using ground penetrating radar},
  journal = {Automation in Construction},
  volume  = {118},
  pages   = {103279},
  year    = {2020},
  doi     = {10.1016/j.autcon.2020.103279}
}

@article{Luo2020,
  author  = {Luo, Tess X. H. and Lai, Wallace W. L.},
  title   = {{GPR} pattern recognition of shallow subsurface air voids},
  journal = {Tunnelling and Underground Space Technology},
  volume  = {99},
  pages   = {103355},
  year    = {2020},
  doi     = {10.1016/j.tust.2020.103355}
}

@article{Hu2023,
  author  = {Hu, Haobang and Fang, Hongyuan and Wang, Niannian and Ma, Duo and
             Dong, Jiaxiu and Li, Bin and Di, Danyang and Zheng, Hongbiao and
             Wu, Jiang},
  title   = {Defects identification and location of underground space for
             ground penetrating radar based on deep learning},
  journal = {Tunnelling and Underground Space Technology},
  volume  = {140},
  pages   = {105278},
  year    = {2023},
  doi     = {10.1016/j.tust.2023.105278}
}

@article{Yang2023,
  author  = {Yang, Xiuwei and Liu, Pingan and Wang, Shujie and Wu, Biyuan and
             Zhang, Kaihua and Yang, Bing and Wu, Xiaohu},
  title   = {Defect detection of composite material terahertz image based on
             {Faster} region-convolutional neural networks},
  journal = {Materials},
  volume  = {16},
  number  = {1},
  pages   = {317},
  year    = {2023},
  doi     = {10.3390/ma16010317}
}

@article{Gao2025,
  author  = {Gao, Mingyu and Huo, Liang and Wang, Fei and Song, Peng and
             Gao, Yulong and Yang, Guohui and Liu, Junyan and Liang, Zhipeng and
             Xie, Yunji and Song, Yinghao},
  title   = {{CNN}-based similar microwave reflection signals for improved
             detectability and intelligent characterization of internal defects
             in composite materials},
  journal = {Journal of Nondestructive Evaluation},
  volume  = {44},
  number  = {1},
  pages   = {28},
  year    = {2025},
  doi     = {10.1007/s10921-025-01163-3}
}

@inproceedings{Yeo2016,
  author    = {Yeo, Hui-Shyong and Flamich, Gergely and Schrempf, Patrick and
               Harris-Birtill, David and Quigley, Aaron},
  title     = {{RadarCat}: Radar categorization for input \& interaction},
  booktitle = {Proceedings of the 29th Annual ACM Symposium on User Interface
               Software and Technology (UIST '16)},
  publisher = {ACM},
  pages     = {833--841},
  year      = {2016},
  doi       = {10.1145/2984511.2984515}
}

@inproceedings{Wang2016,
  author    = {Wang, Saiwen and Song, Jie and Lien, Jaime and Poupyrev, Ivan and
               Hilliges, Otmar},
  title     = {Interacting with {Soli}: Exploring fine-grained dynamic gesture
               recognition in the radio-frequency spectrum},
  booktitle = {Proceedings of the 29th Annual ACM Symposium on User Interface
               Software and Technology (UIST '16)},
  publisher = {ACM},
  pages     = {851--860},
  year      = {2016},
  doi       = {10.1145/2984511.2984565}
}

@inproceedings{Wang2017,
  author    = {Wang, Ju and Xiong, Jie and Chen, Xiaojiang and Jiang, Hongbo and
               Balan, Rajesh Krishna and Fang, Dingyi},
  title     = {{TagScan}: Simultaneous target imaging and material
               identification with commodity {RFID} devices},
  booktitle = {Proceedings of the 23rd Annual International Conference on Mobile
               Computing and Networking (MobiCom '17)},
  publisher = {ACM},
  pages     = {288--300},
  year      = {2017},
  doi       = {10.1145/3117811.3117830}
}

@inproceedings{Dhekne2018,
  author    = {Dhekne, Ashutosh and Gowda, Mahanth and Zhao, Yixuan and
               Hassanieh, Haitham and Choudhury, Romit Roy},
  title     = {{LiquID}: A wireless liquid {ID}entifier},
  booktitle = {Proceedings of the 16th Annual International Conference on Mobile
               Systems, Applications, and Services (MobiSys '18)},
  publisher = {ACM},
  pages     = {442--454},
  year      = {2018},
  doi       = {10.1145/3210240.3210345}
}

@inproceedings{Xie2019,
  author    = {Xie, Binbin and Xiong, Jie and Chen, Xiaojiang and Chai, Eugene and
               Li, Liyao and Tang, Zhanyong and Fang, Dingyi},
  title     = {{Tagtag}: Material sensing with commodity {RFID}},
  booktitle = {Proceedings of the 17th Conference on Embedded Networked Sensor
               Systems (SenSys '19)},
  publisher = {ACM},
  pages     = {338--350},
  year      = {2019},
  doi       = {10.1145/3356250.3360027}
}

@inproceedings{choi2018short,
  title={Short-term load forecasting based on ResNet and LSTM},
  author={Choi, Hyungeun and Ryu, Seunghyoung and Kim, Hongseok},
  booktitle={2018 IEEE international conference on communications, control, and computing technologies for smart grids (SmartGridComm)},
  pages={1--6},
  year={2018},
  organization={IEEE}
}

@ARTICLE{726791,
  author={Lecun, Y. and Bottou, L. and Bengio, Y. and Haffner, P.},
  journal={Proceedings of the IEEE}, 
  title={Gradient-based learning applied to document recognition}, 
  year={1998},
  volume={86},
  number={11},
  pages={2278-2324},
  doi={10.1109/5.726791}}

@inproceedings{konishi2023insar,
  title={INSAR Phase Filtering By Attention-Based Reservoir Computing For High Reliability},
  author={Konishi, Bungo and Hirose, Akira and Natsuaki, Ryo},
  booktitle={IGARSS 2023-2023 IEEE International Geoscience and Remote Sensing Symposium},
  pages={5198--5201},
  year={2023},
  organization={IEEE}
}

\end{document}